\documentclass[reprint, amsmath,amssymb,showpacs,floatfix,longbibliography, aps, twocolumn, superscriptaddress,nofootinbib]{revtex4-1}
\usepackage{times} 
\usepackage{graphicx}
\usepackage{dcolumn}
\usepackage{bm}
\usepackage{color}
\usepackage{xcolor}
\usepackage{hyperref}
\usepackage{enumitem}
\usepackage{cancel}
\usepackage{algorithm} 
\usepackage{algpseudocode} 
\usepackage{tikz}
\usepackage{graphicx}
\usepackage{comment}
\hypersetup{colorlinks=true,citecolor=blue,linkcolor=blue, urlcolor=blue}
\hypersetup{linktocpage}
\newcolumntype{M}[1]{>{\centering\arraybackslash}m{#1}}
\newcolumntype{N}{@{}m{0pt}@{}}
\usepackage{environ}
\usepackage{MnSymbol}

\bibpunct{[}{]}{,}{n}{}{}

\usepackage{amsfonts,amssymb,amsmath}
\usepackage[T1]{fontenc}
\usepackage{tikz}
\usetikzlibrary{arrows,decorations,backgrounds}

\newcommand{\A}{{\mathcal A}}
\newcommand{\B}{{\mathcal B}}

\newcommand{\RR}{{\mathbb R}}

\newcommand{\HH}{{\mathcal{H}}}

\newcommand{\nq}{{L}}  
\renewcommand{\tt}{{t}}

\newcommand{\hc}{{\mathrm{h.c.}}}

\newcommand{\ket}[1]{ | {#1} \rangle}

\let\originalleft\left
\let\originalright\right
\renewcommand{\left}{\mathopen{}\mathclose\bgroup\originalleft}
\renewcommand{\right}{\aftergroup\egroup\originalright}

\newcommand{\da}{\delta_A}
\newcommand{\db}{\delta_B}
\newcommand{\ra}{\rightarrow}
\newcommand{\eps}{\epsilon}

\NewEnviron{eqs}{%
\begin{equation}\begin{split}
    \BODY
\end{split}\end{equation}
}

\begin{document}

\title{Efficient Product Formulas for Commutators and Applications to Quantum Simulation}

\author{Yu-An Chen}
\email[E-mail: ]{yuanchen@umd.edu}
\affiliation {Google Quantum AI, 340 Main Street, Venice, California 90291, USA}
\affiliation{Department of Physics and Joint Quantum Institute, NIST/University of Maryland, College Park, Maryland 20742, USA}
\affiliation{Condensed Matter Theory Center, University of Maryland, College Park, Maryland 20472 USA}
\affiliation{Joint Center for Quantum Information and Computer Science, University of Maryland, College Park,
Maryland 20742, USA}

\author{Andrew M.\ Childs}
\affiliation{Joint Center for Quantum Information and Computer Science, University of Maryland, College Park,
Maryland 20742, USA}
\affiliation {Department of Computer Science, University of Maryland, College Park, Maryland 20742, USA}
\affiliation{Institute for Advanced Computer Studies, University of Maryland, College Park, Maryland 20742, USA}

\author{Mohammad Hafezi}
\affiliation{Department of Physics and Joint Quantum Institute, NIST/University of Maryland, College Park, Maryland 20742, USA}
\affiliation{Joint Center for Quantum Information and Computer Science, University of Maryland, College Park,
Maryland 20742, USA}
\affiliation{Departments of Electrical and Computer Engineering and Institute for Research in Electronics and Applied Physics,
University of Maryland, College Park, Maryland 20742, USA
}

\author{Zhang Jiang}
\email[E-mail: ]{jiangzhang@google.com}
\affiliation {Google Quantum AI, 340 Main Street, Venice, California 90291, USA}

\author{Hwanmun Kim}
\affiliation{Department of Physics and Joint Quantum Institute, NIST/University of Maryland, College Park, Maryland 20742, USA}

\author{Yijia Xu}
\email[E-mail: ]{yijia@umd.edu}
\affiliation{Department of Physics and Joint Quantum Institute, NIST/University of Maryland, College Park, Maryland 20742, USA}
\affiliation{Joint Center for Quantum Information and Computer Science, University of Maryland, College Park,
Maryland 20742, USA}
\affiliation{Institute for Physical Science and Technology, University of Maryland, College Park, Maryland 20742, USA}

\date{\today}

\begin{abstract}
We construct product formulas for exponentials of commutators and explore their applications.
First, we directly construct a third-order product formula with six exponentials by solving polynomial equations obtained using the operator differential method. We then derive higher-order product formulas recursively from the third-order formula. We improve over previous recursive constructions, reducing the number of gates required to achieve the same accuracy. In addition, we demonstrate that the constituent linear terms in the commutator can be included at no extra cost. 
As an application, we show how to use the product formulas in a digital protocol for counterdiabatic driving, which increases the fidelity for quantum state preparation. We also discuss applications to quantum simulation of one-dimensional fermion chains with nearest- and next-nearest-neighbor hopping terms, and two-dimensional fractional quantum Hall phases.
\end{abstract}
\maketitle

\tableofcontents

\section{Introduction}

\begin{figure*}
\includegraphics[width=0.95\textwidth]{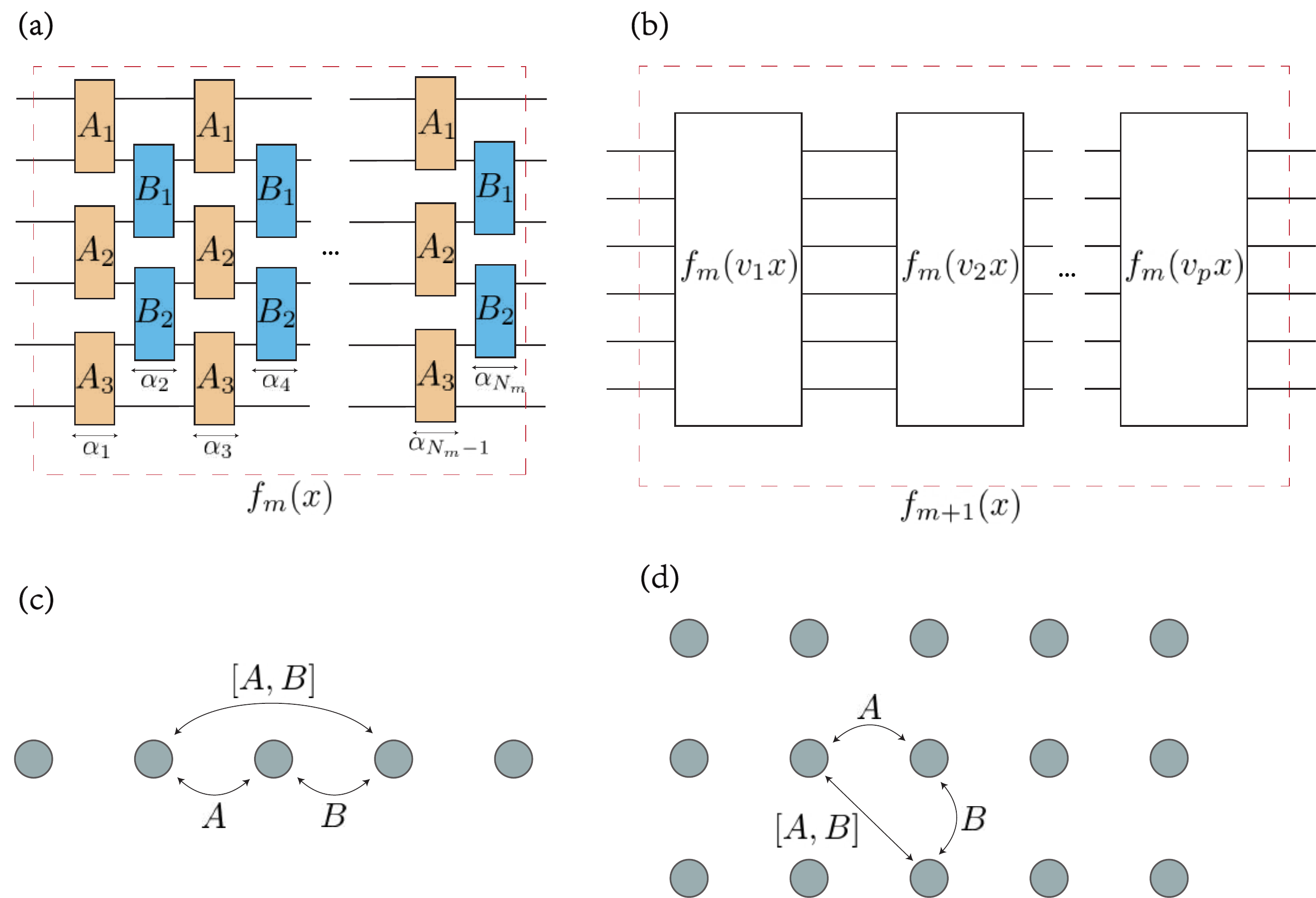}
\caption{(a) Exact construction of a $m$th-order $N_m$-gate product formula. $e^{At}$ and $e^{Bt^\prime}$ are native gates, and the circuit inside the box is the circuit representation of $m$th-order product formula $f_m(x)$. In practice, operator $A = \sum_i A_i$ or $B = \sum_i B_i$ is decomposed as a sum of commuting terms (no overlap between each other). Considering the native gate sets on practical quantum computers, here we use two-qubit gate as an example; (b) Construction of $(m+1)$st-order product formula from $m$th-order formula by $p$-copy recursive formula. The starting point is a $m$th-order product formula $f_m(x)$, and the integrated circuit inside the box is the $(m+1)$st-order product formula $f_{m+1}(x)$, where each component can be $f_m(v_i x)$ or $f_m^{-1} (v_i x)$ (not shown in the figure);
(c)(d) Generate next-nearest-neighbor interaction by the commutator between two nearest-neighbor terms $A$ and $B$ in the 1d chain and the 2d square lattice, e.g.\ Eq.~\eqref{eq: Pauli_next_nearest_neighbor_commutator} and \eqref{eq:next_nearest_neighbor_commutator}.
}
\label{fig: illustration}
\end{figure*}

Product formulas approximate a desired unitary operator with a product of simpler operator exponentials. As a practical tool, product formulas are widely used in quantum simulation of condensed matter models and quantum chemistry problems \cite{Feynman1982,lloyd1996universal,childs2019nearly,tran2020destructive,Childs9456,childs2019theory,Shaw2020_Schwinger_model,Chem_Trotter2015,PNAS_Reiher7555}, as well as quantum Monte Carlo and statistical physics problems \cite{PRA_generalized_Trotter,pre_Trotter_Pathintegral,higher_order_formula_pathintegral,NM05}.
The simplest example is the first-order Lie-Trotter product formula
\begin{eqs}
    e^{xA}e^{xB}=e^{x(A+B)}+O(x^2),
\end{eqs}
which approximates the sum of operators $A$ and $B$. 
From the perspective of quantum simulation, this provides a way to approximate the time evolution of the Hamiltonian $H=A+B$ by multiplying elementary exponentials of the form $e^{-i\delta A}$ and $e^{-i\delta B}$ (which generally do not commute).

In addition to simulating Hamiltonian evolution of a linear combination of terms, one can also construct product formulas for commutators \cite{Jean-Koseleff, Andrew_Nathan}. The simplest product formula for commutators is the second-order formula
\begin{eqs}
    S_2(x):=e^{xA}e^{xB}e^{-xA}e^{-xB}=e^{x^2[A,B]}+O(x^3).
\label{eq: simple product formula for commutator}
\end{eqs}
Such commutator product formulas raise the possibility of simulating complicated unitaries on a quantum simulator using limited native gates: given the time evolution of operators $A$ and $B$, the time evolution of any linear combination of nested commutators involving $A$ and $B$ (i.e., the Lie algebra generated by $A$ and $B$) can be simulated. Moreover, product formulas for commutators with arbitrary high order $k$, $\exp ({[A,B]x^2}) + O(x^{k+1})$, have been constructed recursively \cite{Jean-Koseleff,Andrew_Nathan}.

\begin{table*}[t]
\renewcommand{\arraystretch}{1.25}
\begin{ruledtabular}
\begin{tabular}{ccc}
	&$N_m$-gate $m$th-order product formula & $p$-copy recursive product formula \\
	\hline
	explicit form &$f_m(x)=e^{\alpha_1xA}e^{\alpha_2xB}e^{\alpha_3xA}e^{\alpha_4xB}\ldots e^{\alpha_{N_m}xB}$ & $f_{m+1}(x)=f_m(v_1x)^{\pm 1} f_m(v_2x)^{\pm 1}\ldots f_m(v_px)^{\pm 1}$\\
	elementary unit& native gates $e^{At}, e^{Bt}$& a product formula $f_m(x)$ and inverse formula $f_m(x)^{-1}$\\
	number of gates& $N_m$ & $p N_m - O(p)$ \\
	error & $O(x^{m+1})$ &$O(x^{m+2})$\\
	parameters & $\alpha_1,\alpha_2,\ldots , \alpha_n$ & $v_1,v_2,\ldots , v_p$
\end{tabular}
\end{ruledtabular}
\caption{\label{table: between product and recursive}Comparison between direct and recursive product formulas.}
\end{table*}

Now we introduce terminology for product formulas. An $m$th-order product formula for a sum is a sequence of elementary exponentials of $A$ and $B$ that approximates $\exp\left[x(A+B)\right]$ to $m$th order in $x$:
\begin{eqs}
    e^{t_1A}e^{t_2B}e^{t_3A}e^{t_4B}\ldots =e^{x(A+B)}+O(x^{m+1}),
\end{eqs}
where the time interval for the $i$th elementary exponential is $t_i:=\alpha_ix$ and $\alpha_1,\alpha_2,\alpha_3,\alpha_4,\ldots$ are parameters that define the formula.
Similarly, an $m$th-order commutator product formula is a sequence of elementary exponentials of $A$ and $B$ that approximates $e^{x^2[A,B]}$ to $m$th order:
\begin{eqs}
    e^{t_1A}e^{t_2B}e^{t_3A}e^{t_4B}\ldots =e^{x^2 [A,B]}+O(x^{m+1}).
\label{eq: mth order product formula of commutator}
\end{eqs}
Notice that the prefactor of $[A,B]$ is $x^2$ since it arises from terms quadratic in $t_i = \alpha_i x$.
Both sum and commutator product formulas can be represented as quantum circuits, as shown in Fig.~\ref{fig: illustration}(a).
In general, directly determining suitable coefficients $\alpha_i$ is difficult. Ref.~\cite{NM05} provides an operator differential method for computing the coefficients of commutator product formulas, but it is hard to solve the resulting polynomial equations.

High-order product formulas can be constructed recursively.
In this approach, we choose an invertible product formula $f_m(x)$ as the base formula and recursively increase accuracy with some prescribed sequence of terms of the form $f_m(v_i x)$ and $f_m(v_i x)^{-1}$ for appropriate coefficients $v_i$. By iterating this procedure, we can produce an arbitrarily high-order approximation of the target exponential. Thus the recursive method can be viewed as a ``product formula of product formulas'' where a lower-order product formula is the elementary unit. We call such a recursive formula \emph{$p$-copy} if it uses $p$ elementary formulas $f_m$ and $f_m^{-1}$ to improve the order by 1.
We also consider recursive constructions that use $q$ elementary formulas $f_m$ and $f_m^{-1}$ to improve the order by 2, which we call a $\sqrt{q}$-copy recursive formula.

See Table~\ref{table: between product and recursive} for a comparison of directly constructed product formulas and recursive formulas.
Given an $m$th-order product formula $f_m$ that uses $N_m$ elementary gates, the recursive construction gives am $(m+1)$st-order product formula $f_{m+1}$ with $N_{m+1} = p N_m - O(p)$ gates,\footnote{When we multiply adjacent terms $f_m$ and $f_m^{-1}$ in the recursive formula, two gates in the middle may combine into a single gate if they are both exponentials of the same operator. In the best case, the number of gates is $N_{m+1} = p N_{m}-(p-1)$.} as shown in Fig.~\ref{fig: illustration}(b). Starting from an $m$th-order product formula, we can apply the recursive formula $k$ times to get an $(m+k)$th-order product formula.

\begin{table*}[t]
\renewcommand{\arraystretch}{1.25}
\begin{ruledtabular}
\begin{tabular}{ccccc}
	Recursive formula & Total number of copies & Accuracy improvement & Number of gates \\
	\hline
    Jean-Koseleff \cite[Lemma 7]{Jean-Koseleff}& 3 & $O(x^{m+1})\ra O(x^{m+2})$ & $N_{m+1}=3N_m-2$ \\
    Childs-Wiebe ($5$-copy) \cite[Lemma 7]{Andrew_Nathan} & 5 & $O(x^{m+1}) \ra O(x^{m+2})$ & $N_{m+1}=5N_m-2$\\
    Childs-Wiebe ($\sqrt{6}$-copy) \cite[Thereom 2]{Andrew_Nathan} & 6 & $O(x^{2k+1}) \ra O(x^{2k+3})$ & $N_{2k+2}=6N_{2k}-2$\\
    \hline
    $Q$ ($\sqrt{4}$-copy) &4 & $O(x^{m+1}) \ra O(x^{m+3})$ & $N_{m+2} = 4 N_m - 3$ \\
    $W$ ($\sqrt{5}$-copy) & 5 & $O(x^{m+1}) \ra O(x^{m+3})$ & $N_{m+2}=5N_m-4$\\
    $V$ ($\sqrt{6}$-copy) & 6 & $O(x^{2k}) \ra O(x^{2k+2})$ & $N_{2k+1}=6N_{2k-1}-4$\\
    $\mathcal{G}$ ($\sqrt{10}$-copy) & 10 & $O(x^{m+1}) \ra O(x^{m+3})$ & $N_{m+2}=10N_m-4$
\end{tabular}
\end{ruledtabular}
\caption{\label{table: between recursive formula}Comparison of recursive formulas. The 
first three formulas are previous approaches. The remaining formulas are described in this paper.}
\end{table*}

For practical quantum computation, it is essential to find product formulas that are as efficient as possible. Previous research \cite{Jean-Koseleff,Andrew_Nathan} starts from the second-order commutator product formula Eq.~\eqref{eq: simple product formula for commutator} and uses different recursive methods to improve accuracy. The parameters of these previous recursive constructions are summarized at the top of Table~\ref{table: between recursive formula}.

In general, the number of gates in any $m$th-order product formula is exponential in $m$.\footnote{The number of linearly independent commutators involving $p$ copies of $A$ or $B$ is $\frac{2^p-2}{p}$ if $p$ is a prime \cite{reutenauer93, bourbaki08}. Canceling this many terms generically requires a number of parameters that is exponential in $p$.}
However, distinct methods have different constants that affect the cost of product formulas in practice. In principle, high-order formulas could be constructed directly. However, solving algebraic equations to determine such a formula can be challenging, especially at high orders. Recursive constructions can straightforwardly build higher-order product formulas, potentially at the cost of worse performance than a direct construction.

Finally, we introduce some basic applications of commutator product formulas.
Three-spin interactions can naturally emerge from commutators of two-spin terms. For example,
\begin{eqs}\label{eq: Pauli_next_nearest_neighbor_commutator}
    [\sigma_x^i\sigma_x^j,\sigma_z^j\sigma_z^k]=-2i\sigma_x^i\sigma_y^j\sigma_z^k.
\end{eqs}
More generally, we can construct multi-spin operators from lower order terms.
Furthermore, the commutator between two bosonic or fermionic nearest-neighbor hopping terms is a next-nearest-neighbor hopping term:
\begin{eqs}\label{eq:next_nearest_neighbor_commutator}
    [a_i^\dagger a_j+a_j^\dagger a_i, a_j^\dagger a_k+a_k^\dagger a_j]=a_i^\dagger a_k-a_k^\dagger a_i,
\end{eqs}
where $a_i, a_i^\dagger$ are bosonic or fermionic creation and annihilation operators that obey $[a_i,a_j^\dagger]_\pm=\delta_{i,j}$. This technique can be used to generate complicated interactions from simple nearest-neighbor interactions. In Fig.~\ref{fig: illustration}(c), we show the construction of next-nearest-neighbor hopping in a 1d system using a commutator. We show a similar approach to next-nearest-neighbor hopping for a 2d system in Fig.~\ref{fig: illustration}(d). 

\subsection*{Summary of results}

In this work, we use the operator differential method \cite{NM05} to construct product formulas that combine sums and commutators. Specifically, as described in Sec.~\ref{sec: Third order product formula} and Appendix \ref{sec: operator differential and product formula}, we construct a formula that implements
\begin{eqs}
    e^{ x (A+B) + R x^2 [A,B]} + O(x^4)
\end{eqs}
for arbitrary $R \in \RR$, using 6 exponentials of $A$ and $B$. 

In the large-$R$ limit, our product formula reduces to the pure commutator formula
\begin{align}
    S_3(x) &:=  e^{ \frac{\sqrt{5}-1}{2} x A} e^{ \frac{\sqrt{5}-1}{2} x B}
    e^{ - xA} e^{ -\frac{\sqrt{5}+1}{2} x B} 
    e^{ \frac{3-\sqrt{5}}{2} x A} e^{  x B} \nonumber\\
    &= e^{ x^2 [A,B] } + O(x^4),
\label{eq: third order product formula for commutators}
\end{align}
which reduces the error by one order in $x$ as compared to the group commutator formula Eq.~\eqref{eq: simple product formula for commutator}, giving substantially better performance in practice.

In a recursive construction of higher-order product formulas, a good base formula can have significant impact.
We show that the third-order commutator product formula $S_3(x)$ can improve recursive methods.
We numerically check that previous recursive constructions can perform better when using $S_3(x)$ instead of $S_2(x)$ as the base formula. In particular, this change significantly reduces the total gate count required to achieve a fixed error.

In addition to a better base formula, we improve the recursive method.
We first modify the Childs-Wiebe $\sqrt{6}$-copy formula (Theorem 2 in Ref.~\cite{Andrew_Nathan}), which uses 6 instances of an even-order formula $f_{2k}$ to increase its order by 2. This construction first applies a 2-copy recursive formula to increase the order from $2k$ to $2k+1$ and then applies a 3-copy recursive formula to get a $(2k+2)$nd-order product formula. We observe that these two steps can be decomposed. In particular, if we start with an odd-order product formula, we can apply the 3-copy recursive formula first and then apply the 2-copy one. This modified Childs-Wiebe $\sqrt{6}$-copy formula is denoted $V$ in Table \ref{table: between recursive formula}.

We further propose $\sqrt{4}$-copy, $\sqrt{5}$-copy, and $\sqrt{10}$-copy recursive formulas that use $(4N_m - 3)$, $(5N_m - 4)$, and $(10N_m -4)$ gates, respectively, to generate $(m+2)$nd-order formulas from an $N_m$-gate $m$th-order formula.
Note that using fewer gates to achieve a given order is not necessarily better, since constant factors in the error terms can significantly affect performance.
Indeed, using numerical simulations, we demonstrate that our $\sqrt{10}$-copy recursive formula requires the fewest gates to reach the same accuracy.

In summary, we find that
\begin{enumerate}
    \item The third-order product formula $S_3(x)$ (Eq.~\eqref{eq: third order product formula for commutators}) performs better than the standard choice $S_2(x)$ (Eq.~\eqref{eq: simple product formula for commutator}), serving as a better base formula for all recursive methods in Table~\ref{table: between recursive formula}.
    \item Other recursive formulas can offer better performance, with the $\sqrt{10}$-copy formula performing the best of those we study.
\end{enumerate}

We present two concrete applications of our product formulas. The first application is in the context of the counterdiabatic driving, a method that was originally proposed for analog quantum computation using nested commutators  \cite{CD2003,CD2005,CD_2017,floquet_CD_2019}. In the context of digital quantum state preparation, we demonstrate that our product formulas can generate the commutator terms required for counterdiabatic driving. This additional term increases the fidelity of the final state without increasing the number of gates. To illustrate the efficiency of our approach, we consider state preparation for spins as an example.

The second application is in the context of the quantum simulation. We show that a one-dimensional fermion chain with next-nearest-neighbor hopping terms and a model of two-dimensional fractional quantum Hall phases can be naturally simulated using our method. In both cases, we use nearest-neighbor hopping terms to generate next-nearest-neighbor hopping terms, which break time-reversal symmetry.

The remainder of the paper is organized as follows.
In Sec.~\ref{sec: Third order product formula}, we use the operator differential method to construct the third-order commutator product formula. In addition, we propose a new kind of product formula for combined sums and commutators in Sec.~\ref{sec: sum and commutator}.
Then we derive recursive product formulas for commutators in Sec.~\ref{sec: Recursive relation}. We present numerical simulations of the third-order product formula and recursive constructions in Sec.~\ref{sec: numerical}.
Next, in Sec.~\ref{sec: CD}, we demonstrate that this new formula can implement counterdiabatic driving and generate next-nearest-neighbor interactions from nearest-neighbor terms. We show how to simulate a 1d fermion chain with next-nearest-neighbor hopping terms in Sec.~\ref{sec: 1d fermion chain} and fractional quantum Hall phases (the Kapit-Mueller model) in Sec.~\ref{sec: Kapit-Mueller}. Finally, we discuss open questions and future directions in Sec.~\ref{sec: discussions}.

\section{Third-order product formula} \label{sec: Third order product formula}

In this section, we introduce the third-order exponential product formula, i.e., a formula with error $O(x^4)$ where each elementary exponential has time proportional to $x$. We first present a formula for the pure commutator $[A,B]$ and then present a product formula that includes both the sum $A+B$ and commutator $[A,B]$.

We derive these product formulas using the operator differential method \cite{NM05}, as detailed in Appendix \ref{sec: operator differential and product formula}. In particular, we find the following general expression for a $6$-gate product formula:
\begin{eqs}
    &e^{p_1 x A} e^{p_2 x B} e^{p_3 x A} e^{p_4 x B} e^{p_5 x A} e^{p_6 x B} = \exp ( \Phi (x) )
\end{eqs}
with
\begin{eqs}
    \Phi(x) &= x (lA + mB) + \frac{x^2}{2} (lm-2q) [A,B]  \\
    &\quad+ \frac{x^3}{6}\biggl( \biggl(\frac{l^2 m}{2}-3r\biggr) [A,[A,B]] \\
    &\qquad+ \biggl(\frac{m^2 l }{2} -3s\biggr) [B,[B,A]]\biggr)\\
    &\quad+ O(x^4),
\label{eq: general expression of product formula}
\end{eqs}
where
\begin{eqs}\label{eq: reparameterization}
    l &:= p_1 + p_3 + p_5,  \\ 
    m &:= p_2 + p_4 + p_6,  \\
    q &:= p_2 p_3 + p_2 p_5+ p_4 p_5,  \\
    r &:= p_1 p_2 p_3 + p_1 p_2 p_5 + p_1 p_4 p_5 + p_3 p_4 p_5,  \\
    s &:= p_2 p_3 p_4 + p_2 p_3 p_6 + p_2 p_5 p_6 + p_4 p_5 p_6.
\end{eqs}
Thus an arbitrary $6$-gate product formula $\Phi(x)$ can be reparameterized by $l,m,q,r,s$ which are functions of $p_1,\ldots,p_6$. By fine-tuning the parameters $p_1,\ldots,p_6$ (or equivalently, $l,m,q,r,s$), we can obtain the desired $\Phi(x)$. Here we focus on two cases: a third-order commutator product formula and a third-order product formula for a combination of the sum and commutator.

To construct a third-order product formula for the commutator, we solve for $p_1,\ldots, p_6$ so that
\begin{eqs}\label{eq: conditions_for_pure_commutators}
    l=m&=0,\\
    lm-2q&\neq0,\\
    \frac{l^2m}{2}-3r=\frac{m^2l}{2}-3s&=0.
\end{eqs}
These conditions ensure $\Phi(x)$ only involves the commutator $[A,B]$ and terms that are $O(x^4)$.

Similarly, for the third-order product formula for both sum and commutator, we would like to keep both the linear and (non-nested) commutator terms. Hence we should find $p_1,\ldots, p_6$ so that
\begin{eqs}
    l=m&\neq0,\\
    lm-2q&\neq0,\\
    \frac{l^2m}{2}-3r=\frac{m^2l}{2}-3s&=0.
    \label{eq: conditions_for_sum_and_commutators}
\end{eqs}

We discuss the details in Sec.~\ref{sec: pure commutator} (pure commutator) and Sec.~\ref{sec: sum and commutator} (sum and commutator).

\subsection{Pure commutator}\label{sec: pure commutator}

\begin{figure}
    \centering
    \includegraphics[width=0.45\textwidth]{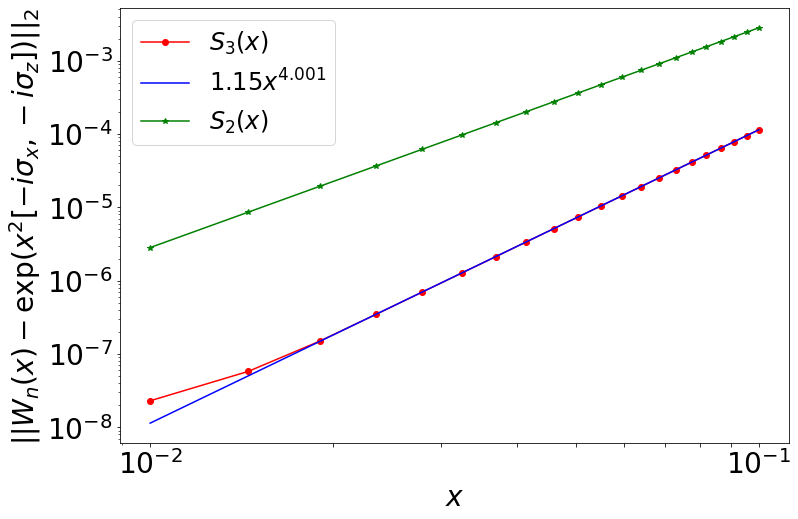}
    \caption{Error scaling for $S_2(x)$ and $S_3(x)$ with $A=-i\sigma_x$, $B=-i\sigma_z$. The error of $S_3(x)$ is smaller than that of $S_2(x)$ by roughly $10^{-2}$. 
    }
    \label{fig:W_3}
\end{figure}

We now derive the $6$-gate third-order product formula for commutators:
\begin{eqs}
    S_3(x) & := \exp \bigl( \tfrac{\sqrt{5}-1}{2} x A\bigr) \exp \bigl( \tfrac{\sqrt{5}-1}{2} x B\bigr) 
    \exp ( - xA) \\
    & \quad \cdot \exp \bigl( -\tfrac{\sqrt{5}+1}{2} x B\bigr) 
    \exp \bigl( \tfrac{3-\sqrt{5}}{2} x A\bigr) \exp ( x B) \\
    &= \exp\bigl( x^2 [A,B] + O(x^4) \bigr).
\label{eq: 6 exp product formula for commutator}
\end{eqs}
This formula can be checked by using the Taylor series of each term up to order $x^3$ to show that the constant, $x$, $x^2$, and $x^3$ terms on both sides agree. In general, by Eq.~\eqref{eq: general expression of product formula}, the product formula
\begin{eqs}
    &e^{p_1 x A} e^{p_2 x B} e^{p_3 x A} e^{p_4 x B} e^{p_5 x A} e^{p_6 x B}\\
    &\quad= \exp ( x^2 [A,B] + O(x^4) )
\end{eqs}
holds if $p_1, p_2, p_3, p_4, p_5, p_6$ satisfy the following polynomial equations resulting from Eq.~\eqref{eq: conditions_for_pure_commutators}:
\begin{eqs}\label{eq: polynomial equation for the 3rd order commutators}
    l=m=r=s&=0,\\
    q= p_2 p_3 + p_2 p_5 + p_4 p_5 &=-1.
\end{eqs}
Eq.~\eqref{eq: 6 exp product formula for commutator} is a particular solution of the above equations. Fig.~\ref{fig:W_3} shows the empirical error scaling behavior of $S_3(x)$ for a $1$-qubit example. The error exponent obtained by fitting the data points in the interval $2\times 10^{-2}\leq x\leq 10^{-1}$ is $4.001$, in good agreement with theory.

\subsection{Sum and commutator}\label{sec: sum and commutator}

In this section, we consider the product formula for both sum and commutator. For arbitrary $R \in \RR$, we want to find a set of parameters $p_1(R),\ldots,p_6(R)$ such that
\begin{equation}
    \Phi(x) = x (A+B) + R ~ x^2 [A, B] + O(x^4).
\end{equation}
This is equivalent to solving the equations given by Eq.~\eqref{eq: conditions_for_sum_and_commutators}:
\begin{eqs}\label{eq:conditions_sum_commutator_compact}
    &l=m=1,\\
    &q=-R+\frac{1}{2},\\
    &r=s=\frac{1}{6}.
\end{eqs}
These equations are in general difficult to solve analytically,
but one can find an approximate solution for large $R$:
\begin{equation}
    \begin{aligned}
    p_1&=(g-1) \sqrt{R + \frac{1}{2}}, & p_2&=(g-1)\sqrt{R + \frac{1}{2}} +1,\\
    p_3 &= -\sqrt{R + \frac{1}{2}} + 1,
    & p_4 &= -g\sqrt{R + \frac{1}{2}},\\
    p_5 &= (2-g) \sqrt{R + \frac{1}{2}},
    & p_6 &= \sqrt{R + \frac{1}{2}} \\
    \end{aligned}
\label{eq:pi}
\end{equation}
where $g:=\frac{\sqrt{5}+1}{2}$. This choice corresponds to
\begin{eqs} \label{eq:approx_soln}
    &l=m=1,\\
    &q= -R +\frac{1}{2},\\
    & r= -(g-1)\biggl(R+\frac{1}{2}-\sqrt{R+\frac{1}{2}}\;\biggr), \\
    & s = (g-1)\biggl(R+\frac{1}{2}-\sqrt{R+\frac{1}{2}}\;\biggr).
\end{eqs}
While this does not satisfy $r=s=\frac{1}{6}$, the leading order of $r$ and $s$ is $O(R)$ (whereas for a general choice of $\{p_i\}$, the leading order is $O(R^{\frac{3}{2}})$).
Substituting Eq.~\eqref{eq:approx_soln} into Eq.~\eqref{eq: general expression of product formula}, we find that the linear term vanishes, the quadratic term is
\begin{eqs}
    \frac{lm-2q}{2} x^2 [A,B] = R x^2 [A,B],
\end{eqs}
and the $x^3$ term is 
\begin{align}
    &\frac{x^3}{6}\biggl( \biggl(\frac{l^2 m}{2}-3r\biggr) [A,[A,B]] + \biggl(\frac{m^2 l }{2} -3s\biggr) [B,[B,A]]\biggr) \nonumber\\
    &= x^3 O(R).
\end{align}
From Eq.~\eqref{eq: general expression of product formula}, we have constructed
\begin{eqs}
    f_R(x)=\exp \bigl[& {x(A+B) + R x^2 [A,B]}  \\
    &+ x^3 O(R)+ x^4 O(R^2) + \cdots \\
    &+ x^k O(R^{\frac{k}{2}}) + \cdots \bigr].
\label{eq: f_R}
\end{eqs}
Notice that the coefficient in front of the $x^k$ term for $k \geq 4$ contains products of $p_{i_1} p_{i_2} \cdots p_{i_k}$, where each $p_{i_j}$ is $O(R^\frac{1}{2})$.

We then use the third-order product formula to implement a new gate from existing gates.
Assume we can perform gates of the form
$e^{\theta_1 A}$ and $e^{\theta_2 B}$ for $\theta_1, \theta_2 \in \RR$
and our goal is to perform
\begin{equation}\label{eq:desired_gate}
e^{\alpha (A+B) + \beta [A,B]}
\end{equation}
for some desired $\alpha,\beta \in \RR$.
First, we pick a large integer $n$ such that $x := \frac{\alpha}{n}$ is small and $R := \frac{\beta n }{\alpha^2}$ is large. Using the function Eq.~\eqref{eq: f_R} constructed above, we have
\begin{eqs}
    f_{R}\left(
    \frac{\alpha}{n}\right)=\exp & \Bigl[
    \frac{\alpha}{n}(A+B) + \frac{\beta}{n}[A,B] + O\Bigl(\frac{\alpha \beta }{n^2}\Bigr)  \\
    &  + \cdots+ O(\frac{\beta^{\frac{k}{2}}}{n^{\frac{k}{2}}} )+ \cdots \Bigr].
\label{eq: product formula}
\end{eqs}
Repeating this function $n$ time gives the desired gate:
\begin{eqs}
    f_{\frac{\beta n}{\alpha^2}}\left(\frac{\alpha}{n}\right)^n=
    & \exp \Bigl[ \alpha (A+B) + \beta [A,B]\\
    &+O\Bigl(\frac{\alpha \beta + \beta^2}{n}\Bigr)\Bigr].
\label{eq: our final form}
\end{eqs}
This implementation uses $6n$ gates and achieves error $O(\frac{1}{n})$. 

In the limit $\alpha \rightarrow 0$, $\beta$ constant, and $n \gg 1$, Eq.~\eqref{eq: product formula} converges to
\begin{align}
    &\exp ( (g-1) \sqrt{\frac{\beta}{n}} A) \exp ( (g-1) \sqrt{\frac{\beta}{n}} B) 
    \exp ( - \sqrt{\frac{\beta}{n}}A) \nonumber\\
    & \quad \cdot \exp ( -g \sqrt{\frac{\beta}{n}} B) 
    \exp ( (2-g) \sqrt{\frac{\beta}{n}} A) \exp ( \sqrt{\frac{\beta}{n}} B) \nonumber\\
    &= \exp\left( \frac{\beta}{n}[A,B] + O(\frac{\beta^2}{n^2} ) \right)
\label{eq: 6 exps commutator}
\end{align}
where $g=\frac{\sqrt{5}+1}{2}$, which reduces to the pure commutator formula in Sec.~\ref{sec: pure commutator}. Notice that the error has the same order with or without the sum $A+B$, which means that there is no extra cost to simulate the sum along with the commutator.

In Sec.~\ref{sec: application}, we use Eq.~\eqref{eq: our final form} to generate new Hamiltonian terms from existing ones, such as next-nearest-neighbor (NNN) hopping terms from nearest-neighbor (NN) hopping terms. There is another useful formula that can easily be derived from Eq.~\eqref{eq: product formula}:
\begin{eqs}
    &\exp \left({\frac{\alpha}{n} C}\right) f_{R}\left(\frac{\alpha}{n}\right)\\
    &=\exp\left(\frac{\alpha}{n}(A+B+C) + \frac{\beta}{n} [A,B] + O\left(\frac{1}{n^2}\right)\right),
\label{eq: modified product formula}
\end{eqs}
or equivalently,
\begin{eqs}
    &\left( \exp \left({\frac{\alpha}{n} C}\right) f_{R}\left(\frac{\alpha}{n}\right) \right)^n \\
    &=\exp\left(\alpha(A+B+C) + \beta [A,B] + O\left(\frac{1}{n}\right)\right),
\end{eqs}
which uses $7n$ gates and has error $O(1/n)$.

\section{Recursive formulas} \label{sec: Recursive relation}

In this section, we introduce the recursive construction of higher-order product formulas. We first focus on the pure commutator formula, where we improve over previous procedures \cite{Jean-Koseleff,Andrew_Nathan}. Then we discuss recursive formulas for both sum and commutator, following the same strategy as the recursive formula for the sum alone \cite{suzuki1990fractal,suzuki_fractal,NM05}. 

\subsection{Pure commutator}

In this section, we introduce recursive formulas that use 4, 5, 6, or 10 copies of an $n$th-order formula to generate an $(n+2)$nd-order formula. To begin, assume we have an $n$th-order formula for the commutator, of the form
\begin{eqs}
    f_n(x) = \exp ( x^2 [A,B]) + C_n x^{n+1}+ D_n x^{n+2} + O(x^{n+3})
\end{eqs}
for some coefficients $C_n,D_n \in R$.
As in previous recursive constructions, we make essential use of inverse product formulas. Given an $n$th-order product formula $f_n(x)$, its inverse formula is $f_n(x)^{-1}$ where 
\begin{eqs}
    f_n(x)^{-1} f_n(x)=1.
\end{eqs}
Since $f_n(x)$ is a product of elementary exponentials
\begin{equation}
    f_n(x)=e^{p_1xA}e^{p_2xB}e^{p_3xA}\ldots e^{p_kxB},
\end{equation}
its inverse is simply
\begin{eqs}
    f_n(x)^{-1}&=\left(e^{p_1xA}e^{p_2xB}e^{p_3xA}\ldots e^{p_kxB}\right)^{-1}\\
    &=e^{-p_kxB}\ldots e^{-p_3xA}e^{-p_2xB}e^{-p_1xA}.
\end{eqs}
Notice that we include coefficients $C_n$ and $D_n$ to keep track of the $x^{n+1}$ and $x^{n+2}$ terms, respectively.
From $f_n(x)$, we can construct other product formulas:
\begin{eqs}
    f_n^{-1}(x) &= \exp ( -x^2 [A,B]) - C_n x^{n+1} - D_n x^{n+2} , \\
    f_n (-x) &= \exp ( x^2 [A,B]) - (-1)^n C_n x^{n+1} \\
    &\quad+ (-1)^n D_n x^{n+2} , \\
    f_n^{-1} (-x) &= \exp ( -x^2 [A,B]) + (-1)^n C_n x^{n+1} \\
    &\quad- (-1)^{n} D_n x^{n+2},
\label{eq: many W}
\end{eqs}
where we omit the $O(x^{n+3})$ error term for brevity. We use $f_n (x)$ and Eq.~\eqref{eq: many W} as building blocks for higher-order product formulas.

In particular, if $n=2k$ is even, there is a recursive formula that increases the order of the product formula by 1 using only 2 copies of the product formula $f_n$ \cite[Corollary 3]{Andrew_Nathan}:
\begin{eqs}\label{eq:2-copy formula}
    f_{2 k+1}(x) &:=  f_{2 k}\left(\frac{x}{\sqrt{2}}\right) f_{2 k}\left(-\frac{x}{\sqrt{2}}\right) \\
    &= \exp \left(x^{2}[A, B]\right)+O\left(x^{2 k+2}\right).
\end{eqs}

For general $n$, there are two previously established ways to increase the order by 1:
\begin{enumerate}
    \item
    The Jean-Koseleff formula \cite{Jean-Koseleff}:
    \begin{eqs}
        f_{n+1}(x)&=\exp(x^2[A,B]+O(x^{n+2}))\\
        &=
        \begin{cases}
            f_n(tx)f_n(sx)f_n(tx) &\text{if $n$ is even}   \\
            f_n(ux)f_n(vx)^{-1}f_n(ux) &\text{if $n$ is odd}  
        \end{cases}
    \label{eq: JK increase 1}
    \end{eqs}
    with $t=(2+2^{2/(n+1)})^{-1/2}$, $s=-2^{1/(n+1)}t$, $u=(2-2^{2/(n+1)})^{-1/2}$, and $v=2^{1/(n+1)}u$.
    \item
    The Childs-Wiebe (5-copy) formula \cite{Andrew_Nathan}:
    \begin{eqs}
        f_{n+1}(x)&=\exp\left(x^2[A,B]+O(x^{n+2})\right) \\
        &=f_{n}(\nu x)^2 f_{n} (\mu x)^{-1} f_{n} (\nu x)^2
    \label{eq: CW increase 1}
    \end{eqs}
    with $ \mu = (4s_n)^{1/2}$, $\nu=(1/4+ \sigma)^{1/2}$, and $\sigma =\frac{4^{\frac{2}{n+1}}}{4(4-4^{\frac{2}{n+1}})}$.
\end{enumerate}

\subsubsection{\texorpdfstring{$\sqrt{6}$}{}-copy recursive formula}

Theorem 2 of Ref.~\cite{Andrew_Nathan} defines a $\sqrt{6}$-copy recursive construction that improves the order of an even-order product formula $f_{2k}(x)$ by 2 using $6$ copies of $f_{2k}(x)$ and $f_{2k}(x)^{-1}$. We observe that this construction first applies the 2-copy formula Eq.~\eqref{eq:2-copy formula} and then applies the Jean-Koseleff formula Eq.~\eqref{eq: JK increase 1}. Alternatively, we can consider these two steps independently and combine them in different ways. In particular, given an odd-order product formula, we can first apply the the Jean-Koseleff formula and then apply the 2-copy formula.

Here we explicitly describe this alternative
$\sqrt{6}$-copy recursion for odd-order product formulas.
Let $V_n(x)$ be a product formula that approximates $\exp(x^2[A,B])$ with error $O(x^{n+1})$ for odd $n$.

The two-step recursive relation has the form
\begin{eqs} \label{eq:6-copy formula}
    V_{n+2}(x)&=V_n\left(\frac{ux}{\sqrt{2}}\right)V_n\left(\frac{vx}{\sqrt{2}}\right)^{-1}V_n\left(\frac{ux}{\sqrt{2}}\right)\\
    &\quad\times V_n\left(\frac{-ux}{\sqrt{2}}\right)V_n\left(\frac{-vx}{\sqrt{2}}\right)^{-1}V_n\left(\frac{-ux}{\sqrt{2}}\right),
\end{eqs}
where $u=(2-2^{2/(n+1)})^{-1/2}$ and $v=2^{1/(n+1)}u$. Here the first three terms and the last three terms are $(n+1)$st-order formulas, which combine to give an $(n+2)$nd-order formula.
This construction increases the order by 2 using 6 copies of $V_n$ and $V_n^{-1}$, so we call it a $\sqrt{6}$-copy recursion.

This approach can applied to our 3rd-order formula $S_3(x)$ to get higher-order formulas. Letting $N^V_n$ denote the number of gates in the $n$th-order formula, we have
\begin{eqs}
    N^V_{n+1}=
    \begin{cases}
        3N^V_n -2 & \text{if $n$ is odd,} \\ 
        2N^V_2 & \text{if $n$ is even.}
    \end{cases}
\end{eqs}
Starting from $N_3 = 6$, this gives
\begin{eqs}
    N^V_{2k+1} = \frac{1}{15}(13 \cdot 6^k+ 12).
\end{eqs}

\subsubsection{\texorpdfstring{$\sqrt{10}$}{}-copy recursive formula}

Let $\mathcal{G}_n(x)$ be an invertible product formula that approximates $\exp(x^2[A,B])$ with error $O(x^{n+1})$. If $n$ is odd, the Childs-Wiebe (5-copy) formula, Eq.~\eqref{eq: CW increase 1}, can be used to increase its order by 1. If $n$ is even, we can apply Eq.~\eqref{eq:2-copy formula} to increase the order by 1 using 2 copies of $\mathcal{G}_n$.
Overall, we use 10 copies of $\mathcal{G}_n$ and $\mathcal{G}_n^{-1}$ to increase the order by 2. Therefore, $\mathcal{G}_n(x)$ is a $\sqrt{10}$-copy product formula.

Let $N^{\mathcal{G}}_n$ denote the number of gates in the $n$th-order formula.
For odd $n=2k+1$, we have
\begin{eqs}
    N^{\mathcal{G}}_{2k+2} = 5 N^{\mathcal{G}}_{2k+1} -2,
\label{eq: 6 copy recursive 1}
\end{eqs}
and for even $n=2k$, we have
\begin{eqs}
    N^{\mathcal{G}}_{2k+1} = 2 N^{\mathcal{G}}_{2k}.
\label{eq: 6 copy recursive 2}
\end{eqs}
Combining Eqs.~\eqref{eq: 6 copy recursive 1} and \eqref{eq: 6 copy recursive 2}, we have
\begin{eqs}
    N^{\mathcal{G}}_{2k+3} = 10N^{\mathcal{G}}_{2k+1} -4.
\end{eqs}
Starting from the base formula with $N_3 = 6$, we have
\begin{eqs}
    N^{\mathcal{G}}_{2k+1} = \frac{1}{9} (5 \cdot 10^{k} + 4). 
\end{eqs}

\subsubsection{\texorpdfstring{$\sqrt{5}$}{}-copy recursive formula}

We now consider the product
\begin{eqs}
    &W_n(-s'x) W_n^{-1}(x) W_n(sx) W_n^{-1}(-x) W_n(-s'x) \\
    &= \exp((s^2 + 2{s'}^2-2) x^2 [A,B]) \\
    &\quad+ (s^{n+1}+2{s'}^{n+1} -2) C_n x^{n+1}\\
    &\quad+(s^{n+2}-2 {s'}^{n+2}) D_n x^{n+2} + O(x^{n+3}).
\end{eqs}
We first choose $s'= 2^{-\frac{1}{n+2}} s$ such that $s^{n+2}-2 {s'}^{n+2}=0$. To eliminate the coefficient $(s^{n+1}+2{s'}^{n+1} -2)$, we have $s^{n+1}+2{s'}^{n+1} = (1+2^{\frac{1}{n+2}}) s^{n+1} = 2$. Therefore, we choose $s=(\frac{2}{1+2^{\frac{1}{n+2}}})^{\frac{1}{n+1}}$. Then we define a new variable
\begin{eqs}
    x' &:= x \sqrt{(s^2 + 2{s'}^2-2)} \\
    & = x \sqrt{\left(\frac{2}{1+2^{\frac{1}{n+2}}}\right)^{\frac{2}{n+1}}(1+2^{\frac{n}{n+2}}) - 2 } \\
    &=: r x.
\end{eqs}
We can check that $r>0$ for $n>1$.
Finally, we get the $(n+2)$nd-order formula
\begin{eqs}
    &W_{n+2}(x')\\
    &=
    W_n(-s' \tfrac{x'}{r})
    W_n^{-1}(\tfrac{x'}{r})
    W_n(s \tfrac{x'}{r})
    W_n^{-1}(-\tfrac{x'}{r})
    W_n(-s' \tfrac{x'}{r}).
\end{eqs}
Letting $N^W_n$ denote the number of gates in $W_n$, we have the recursive relation
\begin{eqs}
    N^W_{n+2} = 5 N^W_{n} -4.
\end{eqs}
With $N_3=6$, we have
\begin{eqs}
    N^W_{2k+1}=5^k+1.
\end{eqs}

\subsubsection{\texorpdfstring{$\sqrt{4}$}{}-copy recursive formula}

\renewcommand{\arraystretch}{1.25} 
\begin{table*}
\caption{\label{sec 3:table_numerical_solution_cd}Numerical solutions of Eq.~\eqref{eq: 4 gates conditions} for $n=3,5,7,9,11$.}
\begin{ruledtabular}
\begin{tabular}{cccccc}
	&$n=3$ & $n=5$ & $n=7$ & $n=9$ & $n=11$\\
	\hline
	$a$ & $1$ & $1$ & $1$ & $1$ & $1$ \\
	$b$ & $2$ & $2$ & $2$ & $2$ & $2$ \\
	$c$ & $1.982590733$ & $1.996950166$ & $1.999411381$ & $1.999880034$ & $1.999974677$\\
	$d$ & $-0.8190978288$ & $-0.8642318466$ & $-0.8911860667$ & $-0.9091844711$ & $-0.9220693131$\\
	$a^2-b^2+c^2-d^2$ & 0.2597447625 & 0.2409130177 & 0.2034332678 & 0.1729037481 & 0.1496868917\\
\end{tabular}
\end{ruledtabular}
\end{table*}

We now discuss a way to use only 4 copies of an $n$th-order product formula to generate an $(n+2)$nd-order product formula. Let $Q_n(x)$ be an invertible product formula that approximates $\exp(x^2[A,B])$ with error $O(x^{n+1})$. Consider the following product:
\begin{eqs}
    &Q_n(a x) Q_n^{-1}(b x) Q_n (c x ) Q_n^{-1}(d x) \\
    =& \exp((a^2-b^2+c^2-d^2) x^2 [A,B]) \\
    &+(a^{n+1}-b^{n+1}+c^{n+1}-d^{n+1})C_n x^{n+1} \\
    &+ (a^{n+2} - b^{n+2} + c^{n+2} - d^{n+2}) D_n x^{n+2} + O(x^{n+3}).
\end{eqs}
To produce a formula of order $n+2$, we want to find $a,b,c,d$ satifying
\begin{eqs}
    a^{n+1}-b^{n+1}+c^{n+1}-d^{n+1} &= 0 \\
    a^{n+2}-b^{n+2}+c^{n+2}-d^{n+2} &= 0 \\
    a^2-b^2+c^2-d^2 & \neq 0.
\label{eq: 4 gates conditions}
\end{eqs}
If a solution exists, we can define a new variable $x' = s x$, with $s := \sqrt{|a^2-b^2+c^2-d^2|}$, to find the $(n+2)$nd-order formula
\begin{eqs}
    & Q_n(\tfrac{a}{s} x') Q_n^{-1}(\tfrac{b}{s} x') Q_n (\tfrac{c}{s} x' ) Q_n^{-1}(\tfrac{d}{s} x') \\
    &=
\begin{cases}
    Q_{n+2} (x') & \text{if } a^2-b^2+c^2-d^2 > 0\\
    Q^{-1}_{n+2} (x')  &\text{if } a^2-b^2+c^2-d^2 < 0.
\end{cases}
\end{eqs}
Let $N^Q_n$ be the number of gates of the $n$th-order formula.
The number of gates in such a formula satisfies
\begin{eqs}
    N^Q_{2k+3} = 4 N^Q_{2k+1} - 3.
\end{eqs}
With $N_3=6$, we have
\begin{eqs}
    N^Q_{2k+1}=5 \cdot 4^{k-1}+1.
\end{eqs}
We can take $a=1$, $b=2$, and numerically solve Eq.~\eqref{eq: 4 gates conditions} to find $c$ and $d$. Table \ref{sec 3:table_numerical_solution_cd} presents numerical solutions for $n=3,5,7,9,11$. We prove in Appendix \ref{sec: existence of solution} that a solution exists for general $n$.

\subsection{Sum and commutator}\label{Sec 3:recursive_for_sum_commutators}

We also construct a recursive formula for the product formula Eq.~\eqref{eq: product formula} for a linear combination of a sum and a commutator, using the same idea as the $\sqrt{6}$-copy approach described above. Suppose we have an $m$th-order product formula of the form
\begin{align}
    f_{R,m}(x)&=\exp\left(x(A+B)+\frac{x\beta}{\alpha}[A,B]\right)+C_{m}x^{m+1} \nonumber\\
    &\quad+O(x^{n+2}).
\end{align}
In this formula, the commutator term scales with $x$ instead of $x^2$. While the commutator term is $x^2 R [A,B]$ in Eq.~\eqref{eq: f_R}, it becomes $\frac{x\beta}{\alpha}[A,B]$ with the choice of large $R=\frac{\beta}{\alpha x}$.
Since $A+B$ and $[A,B]$ are both first-order terms, the recursive formula for sum and commutator should be similar to the recursive formula for sum. Here we use Suzuki's method \cite{suzuki_fractal} to construct the recursive product formula for sum and commutator.

If $m$ is even, then we consider the $3$-copy sequence
\begin{eqs}
    &f_{R,m}(ax)f^{-1}_{R,m}(bx)f_{R,m}(ax)\\
    &=\exp\left((2a-b)x(A+B)+(2a-b)\frac{x\beta}{\alpha}[A,B]\right)\\
    &\quad+(2a^{m+1}-b^{m+1})C_{m}x^{m+1}+O(x^{n+2}).
\end{eqs}
To obtain an $(m+1)$st-order product formula, $a, b$ should satisfy
\begin{eqs}
    2a-b=1,\quad 2a^{m+1}-b^{m+1}=0
\end{eqs}
to eliminate the $(2a^{m+1}-b^{m+1})C_{m+1}x^{m+1}$ term. The solution is
\begin{equation}
    a=(2-2^{1/(m+1)})^{-1}, \quad b=2^{1/(m+1)}a.
\end{equation}
If $m$ is odd, then we consider the $2$-copy sequence
\begin{eqs}
    &f_{R,m}(-ax)^{-1}f_{R,m}(bx)\\
    &=\exp\left((a+b)x(A+B)+(a+b)\frac{x\beta}{\alpha}[A,B]\right)\\
    &\quad+(-a^{m+1}+b^{m+1})C_{m}x^{m+1}+O(x^{n+2}).
\end{eqs}
To eliminate the $C_{m}x^{n+1}$ term, we must have
\begin{eqs}
    a+b=1,\quad -a^{m+1}+b^{m+1}=0,
\end{eqs}
which is satisfied with
\begin{eqs}
    a=b=\frac{1}{2}.
\end{eqs}

\section{Numerical evidence}\label{sec: numerical}

\begin{figure}
    \centering
    \includegraphics[width=0.45\textwidth]{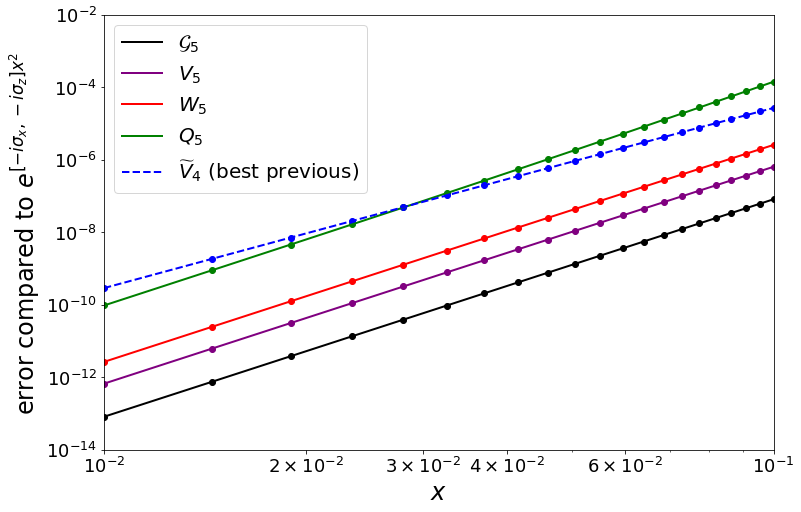}
    \caption{Error scaling for $\mathcal{G}_5$, $V_5$, $W_5$, $Q_5$ and the previous best formula $\widetilde{V}_4$. We use the spectral norm $\|f(x)-\exp([-i\sigma_x,-i\sigma_z]x^2)\|$ to measure the error.
    By fitting the last 10 points for each formula ($0.05\leq x\leq 0.1$), we find the slopes for $\mathcal{G}_5$, $V_5$, $W_5$, $Q_5$ and $\widetilde{V}_4$ are 6.001, 5.958, 5.967, 6.371, and 4.920, respectively.
    }
    \label{fig:W_5}
\end{figure}

The analytical formulas presented above indicate how the errors scale with powers of $x$. However, the constant factors in the error terms of different product formulas significantly affect their performance in practice. To better understand this, we numerically compare the different approaches. Specifically, we evaluate the performance of our $\sqrt{10}$-copy, $\sqrt{6}$-copy, $\sqrt{5}$-copy, and $\sqrt{4}$-copy recursive formulas built from the base formula $S_3(x)$, and compare them with the previous best method, the Childs-Wiebe $\sqrt{6}$-copy formula \cite{Andrew_Nathan} (which is built from the base formula $S_2(x)$).

We evaluate the commutator product formulas with $A=-i\sigma_x$ and $B= -i \sigma_z$ for $x \in [10^{-2}, 10^{-1}]$. Fig.~\ref{fig:W_5} plots the error compared to the exact exponential of the commutator $[A,B]$. The errors scale as $x^{6.001}$ ($\mathcal{G}_5$), $x^{5.958}$ ($V_5$), $x^{5.867}$ ($W_5$), $x^{6.371}$ ($Q_5$), and $x^{4.920}$ ($\widetilde{V}_4$), in good agreement with the analytical scalings.

Next, we compare the number of gates required to achieve a fixed accuracy for different recursive formulas.

We set $e^{-i \sigma_x x}$, $e^{-i \sigma_z x}$ as our elementary exponentials and $\exp([-i\sigma_x x,-i\sigma_z x])$ as our target. We numerically determine the minimum number of elementary exponentials (i.e., gates) to achieve a fixed accuracy $\|f(x)-\exp([-i\sigma_xx,-i\sigma_zx])\|_2=10^{-4}$ for different approximation formulas $f(x)$.

\begin{figure}
    \centering
    \includegraphics[width=0.45\textwidth]{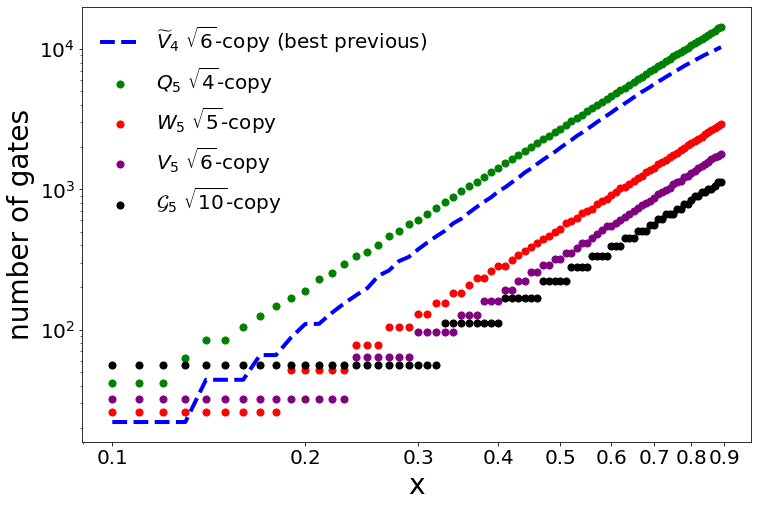}
    \caption{Number of gates to achieve $\exp (x^2 [-i \sigma_x, -i \sigma_z])$ within error $10^{-4}$.}
\label{2:error_diff_formula}
\end{figure}

\begin{figure}
    \centering
    \includegraphics[width=0.45\textwidth]{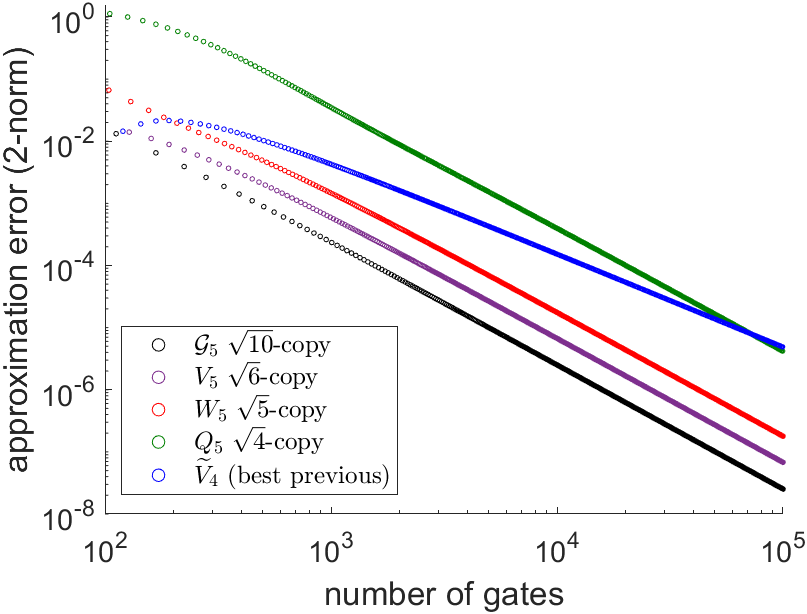}
    \caption{Simulation error of $\exp([-i\sigma_x,-i\sigma_z])$ for different gate numbers and formulas.}
    \label{fig:error_diff_gates}
\end{figure}

We calculate the number of gates to achieve error at most $10^{-4}$ using the aforementioned product formulas. Specifically, we consider 5th-order product formulas obtained from the base formula $S_3$ using the $\sqrt{4}$-copy ($Q_5$), $\sqrt{5}$-copy ($W_5$), $\sqrt{6}$-copy ($V_5$) and $\sqrt{10}$-copy ($\mathcal{G}_5$) approaches. We compare them with the $4$th-order formula $\widetilde{V}_4$ obtained from the base formula $S_2$ using the Childs-Wiebe $\sqrt{6}$-copy recursion \cite{Andrew_Nathan}. See Fig.~\ref{2:error_diff_formula} for the numerical comparison. The number of gates for the $\sqrt{10}$-copy formula $\mathcal{G}_5$ is constant in the interval $x\in[0.1,0.3]$ since its error is always below the threshold. Asymptotically, the scaling of the number of gates for an $n$th-order formula to achieve fixed accuracy is $O(x^{\frac{2n+2}{n-1}})$.
Fig.~\ref{2:error_diff_formula} numerically shows that the $\sqrt{10}$-copy formula has the best performance. Although the $\sqrt{4}$-copy, $\sqrt{5}$-copy, $\sqrt{6}$-copy, and $\sqrt{10}$-copy formulas all have the same error scaling, the constant factors determine their performance in practice.

Fig.~\ref{fig:error_diff_gates} shows the error in simulating $\exp([-i\sigma_x,-i\sigma_z])$ using $\widetilde{V}_4$ (the best previous method), $Q_5$, $W_5$, $V_5$, and $\mathcal{G}_5$. The horizontal axis indicates the total number of elementary exponentials, while the vertical axis indicates the simulation error $\|f(1/\sqrt{r})^r-\exp([-i\sigma_x,-i\sigma_z])\|_2$, where $r$ is the number of time steps used in the simulation and $f(x)$ is the product formula. In an $r$-step simulation, the total number of elementary exponentials for $Q_5$, $W_5$, $V_5$, $\mathcal{G}_5$ and $\widetilde{V}_4$ are $21r$, $26r$, $32r$, $56r$, and $22r$, respectively. The numerical results show that the larger number of exponentials in each time step of $V_5$ and $\mathcal{G}_5$ is offset by their reduced error.

Fig.~\ref{2:error_diff_formula} and Fig.~\ref{fig:error_diff_gates} show that $V_5$ and $\mathcal{G}_5$ improve upon the best previous result $\widetilde{V}_4$. Hybrid approaches that combine previous recursive formulas with our new base formula $S_3(x)$ also give improvements over $\widetilde{V}_4$, but they do not perform as well as $W_5$, $V_5$, and $\mathcal{G}_5$, so we do not include them in Fig.~\ref{2:error_diff_formula} and Fig.~\ref{fig:error_diff_gates}.

\section{Applications to quantum simulation} \label{sec: application}

\subsection{Counterdiabatic driving} \label{sec: CD}

In this section, we discuss using commutator product formulas to implement counterdiabatic driving (CD) \cite{CD2003, CD2005, Berry2009, CD_2017}. In an adiabatic process $\HH(\lambda(t))$, the time evolution $\exp( - i \int \mathrm{d}t \, \HH(\lambda(t)) )$ keeps the system in its instantaneous ground state if $\lambda(t)$ is slowing varying. In other words,
\begin{eqs}
    \ket{\Psi (\tau)} \approx \exp\left( -i \int_0^\tau \mathrm{d}t \, \HH(\lambda(t)) \right) \ket{\Psi (0)} ,
\end{eqs}
where $\ket{\Psi (t)}$ denotes the ground state of $\HH(\lambda(t))$. In general, this approximation fails if $\lambda(t)$ varies too rapidly. However, by introducing counterdiabatic driving terms, the system can remain in the ground state even though $\lambda(t)$ varies rapidly. Specifically,
\begin{eqs}
    \ket{\Psi (\tau)} \approx \exp\left( -i \int_0^\tau \mathrm{d}t \, \HH_{\mathrm{CD}}(\lambda(t)) \right) \ket{\Psi (0)}
\end{eqs}
where \cite{Berry2009, floquet_CD_2019}
\begin{eqs}
    \HH_{\mathrm{CD}}(\lambda(t)) = \HH (\lambda(t)) + \dot{\lambda} C_\lambda
\end{eqs}
with $\langle m | C_\lambda | n \rangle = -i\frac{\langle m | \partial_\lambda \HH | n \rangle }{\epsilon_m - \epsilon_n}$, where $|n \rangle$ denotes an eigenstate of $\HH (\lambda)$ with energy $\epsilon_n$ , i.e., $\HH (\lambda) \ket{n} = \epsilon_n \ket{n}$.

Ref. \cite{floquet_CD_2019} proposes using Floquet engineering to generate the CD term $C_\lambda$.
This term can be expressed as the sum of nested commutators \cite{floquet_CD_2019}
\begin{eqs}
    C_\lambda = i \sum_k c_k (\lambda) \underbrace{[H,[H, \dots [H}_{2k-1} , \partial_\lambda H]]],
\end{eqs}
where the coefficients $c_k(\lambda)$ are determined by minimizing the action
\begin{eqs}
    S_l = \text{Tr}[G_l^2]
\end{eqs}
with
\begin{eqs}
    G_l = \partial_\lambda \HH - i[\HH, C_\lambda].
\end{eqs}
For simplicity, we truncate to only the first term, giving
\begin{eqs}
    C_\lambda \approx i c_1(\lambda) [\HH, \partial_\lambda \HH].
\end{eqs}

Commutator product formulas can be used to implement the CD term. To demonstrate this application, we consider the case $H(\lambda) = H_0 + \lambda H_1$. The time evolution is
\begin{eqs}
    &\exp (- i\int \mathrm{d}t \, \HH_{\mathrm{CD}} (t)) \\
    &= \exp \left(-\int \mathrm{d}t \, (i \HH_0 + i \lambda \HH_1 - \Dot{\lambda} c_1 (\lambda) [\HH_0, \HH_1]) \right).
\end{eqs}
For each infinitesimal time interval $[t, t+ \delta t]$, we apply the following unitary operator:
\begin{eqs}
    \exp(-i \HH_0 \delta t - i \lambda \HH_1 \delta t + \Dot{\lambda} c_1 (\lambda) [\HH_0, \HH_1] \delta t).
\end{eqs}
This unitary operator can be simulated by the product formula Eq.~\eqref{eq: product formula} with $A= -i \HH_1$, $B= -i \lambda \HH_0$, $n=\frac{1}{\delta t}$, $\alpha = 1$, and $\beta = \frac{\Dot{\lambda} c(\lambda)}{\lambda}$. The product-formula error is $O( (\beta + \beta^2) \delta t)$.

As a concrete example, consider using the product formula to simulate the counterdiabatic time evolution of the time-dependent two-qubit Hamiltonian $\HH(\lambda)= \HH_A + \HH_B$ where
\begin{eqs}
    \HH_A &:= h_z(\lambda-1)(\sigma_1^z+\sigma_2^z) \\
    \HH_B &:= J(\sigma_1^x\sigma_2^x+\sigma_1^z\sigma_2^z)
\end{eqs}
with
$\lambda(t)=\sin^2\left(\frac{\pi}{2}\sin^2\left(\frac{\pi t}{2\tau}\right)\right)$. Notice that $\HH_A$ and $\HH_B$ do not commute. The first-order counterdiabatic driving term is \cite{floquet_CD_2019}
\begin{eqs}\label{eq:2-qubit_CD}
    C_\lambda&=-\frac{Jh_z}{2}\frac{(\sigma_1^y\sigma_2^x+\sigma_1^x\sigma_2^y)}{J^2+4(\lambda-1)^2h_z^2}. 
\end{eqs}
We can write Eq.~\eqref{eq:2-qubit_CD} as a commutator between $\HH_A$ and $\HH_B$:
\begin{eqs}\label{eq: CD_commutator}
    C_\lambda&=i\frac{1}{4(1-\lambda)}\frac{[-i\HH_A,-i\HH_B]}{J^2+4(\lambda-1)^2h_z^2}.
\end{eqs}
Hence we can construct the first-order counterdiabatic term using a commutator product formula. In the product formula setting, we choose $A=-i\HH_A$, $B=-i\HH_B$, $n= \frac{1}{\delta t}$, $\alpha=1$, and $\beta=\dot{\lambda}\frac{1}{4(1-\lambda)}\frac{1}{J^2+4(\lambda-1)^2h_z^2}$ to implement the counterdiabatic Hamiltonian over a time interval $\delta t$. Defining $R = \frac{\beta n}{\alpha^2}$, the product formula Eq.~\eqref{eq: product formula} gives
\begin{eqs}
    f_{R}\left( \delta t\right)&=
    e^{p_1(R) A \delta t} e^{p_2(R) B \delta t} e^{p_3(R) A \delta t} \\
    &\quad \cdot e^{p_4(R) B\delta t} e^{p_5(R) A\delta t} e^{p_6(R) B\delta t} \\
    &=\exp  \Big( \delta t(A+B) + \beta \delta t[A,B] + O(\delta t^2)\Big) \\
    &= \exp \Big( -i (\HH_A + \HH_B) \delta t -i (\dot{\lambda} C_\lambda) \delta t\Big) \\
    &= \exp ( -i \HH_{\mathrm{CD}} \delta t ),
\end{eqs}
where $p_i(R)$ is the solution Eq.~\eqref{eq:pi}, which provides a good approximation provided $R$ is large (i.e., $\delta t$ is small).\footnote{For arbitrary $R$, the solution $p_i (R)$ can be determined numerically using Eq.~\eqref{eq:conditions_sum_commutator_compact}. However, here we choose $\delta t$ to be small so that the solution $p_i(R)$ has the convenient form of Eq.~\eqref{eq:pi}} The overall counterdiabatic time evolution is the product of $f_{R}(\delta t)$ for each time interval, i.e.,
\begin{eqs}
    \exp( -i \int_0^\tau \mathrm{d}t \, \HH_{\mathrm{CD}}(\lambda(t)) ) = \prod_{k=0}^{N-1} f_{R(t_k)} (\delta t)
\label{eq: CD products}
\end{eqs}
where $t_k = \frac{k}{N}\tau$ and $\delta t = \frac{\tau}{N}$. Notice that in each term, the operator $A$ and the parameter $R$ depend on $t_k$.

For comparison, the time evolution of the original Hamiltonian can be simulated as
\begin{eqs}
    \exp\left( -i \int_0^\tau \!\! \mathrm{d}t \, \HH(\lambda(t)) \right) = \prod_{k=0}^{N-1} (e^{-i \HH_A(t_k) \delta t /3}e^{-i \HH_B \delta t /3})^3,
\label{eq: Trotter products}
\end{eqs}
where we use first-order Trotterization, $e^{A \delta t/3} e^{B \delta t/3} = e^{(A+B) \delta t/3} + O (\delta t^2)$. We choose $\delta t /3$ in each time step to match the total gate number in the counterdiabatic simulation, which uses 6 exponentials for each $t_k$.

\begin{figure}
    \centering
    \includegraphics[width=0.45\textwidth]{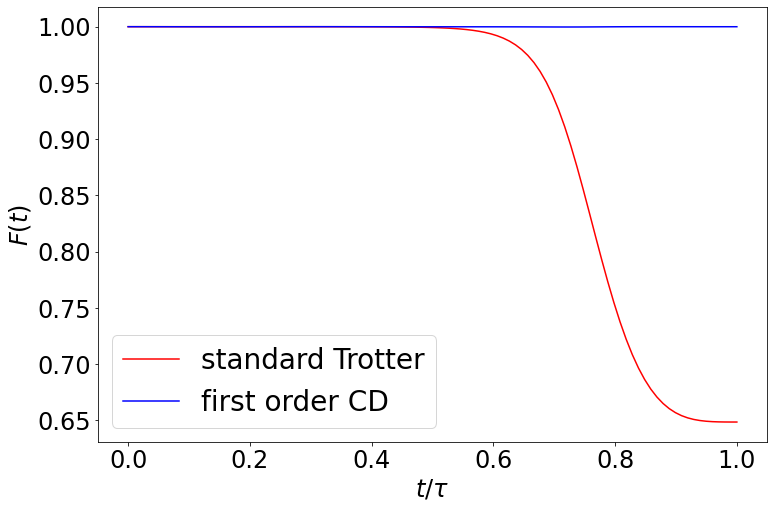}
    \caption{ The fidelity of the evolved state under standard Trotterization and CD protocols. The Hamiltonian has $J=-1$ and $h_z=5$, the total evolution time is $\tau=1$, and the number of steps is $N=100$. There are 6 exponentials in each step, so each simulation uses 600 gates in total. The CD protocol remains close to the ground state, while the standard Trotterization approach starts to deviate from the ground state after $t \approx 0.6 \tau$.}
    \label{fig:my_label}
\end{figure}

\begin{figure}
    \centering
    \includegraphics[width=0.45\textwidth]{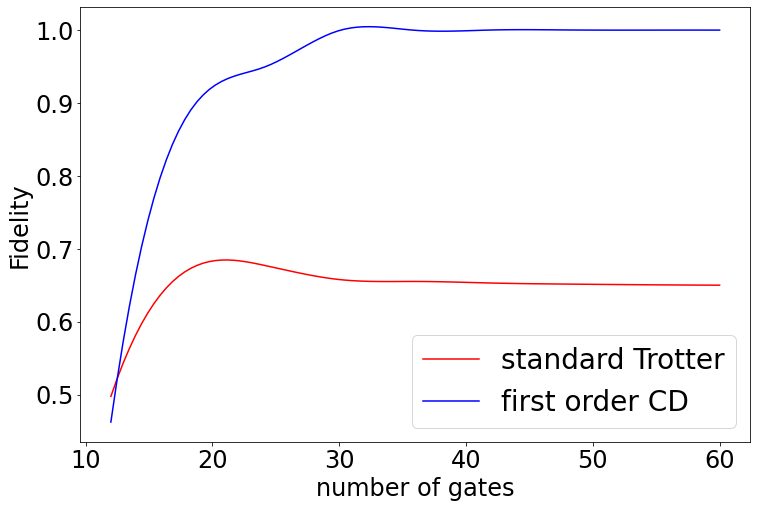}
    \caption{Comparison of the final fidelity at $t=\tau$ using the standard Trotter protocol and counterdiabatic driving approach.}
    \label{fig:digital_CD}
\end{figure}

We consider approximations to the evolution from $t=0$ to $t'=r\delta t$ of $\HH(\lambda(t))$ and $\HH_{\text{CD}}(\lambda(t))$ by standard Trotterization and our digital CD approach, respectively. 

The evolved states with these approximations are
\begin{eqs}
    \ket{\Psi^{\text{evolved}}_{\text{Trotter}} (t')} 
    &= \prod_{k=0}^{r-1} (e^{-i \HH_A(t_k) \delta t /3}e^{-i \HH_B \delta t /3})^3 \ket{\Psi(0)}, \\
    \ket{\Psi^{\text{evolved}}_{\text{CD}}(t')}
    &= \prod_{k=0}^{r-1} f_{R(t_k)} (\delta t) \ket{\Psi(0)}.
\end{eqs}
We define the fidelity of the process as the overlap with the ground state $\ket{\Psi (t')}$ of $\HH(\lambda(t))$:
\begin{eqs}
    F_\alpha (t') := | \langle \Psi (t')  \ket{\Psi^{\text{evolved}}_\alpha (t')} |^2, ~ \alpha \in \{\mathrm{Trotter}, \mathrm{CD}\}.
\end{eqs}

Fig.~\ref{fig:my_label} shows a numerical computation of these quantities. We see that the product formula simulation of counterdiabatic evolution uses the commutator term to keep the system close to the ground state, while the corresponding evolution with standard Trotterization deviates from the ground state after $t \approx 0.6 \tau$.

We also examine the performance of the standard Trotter method and our digital CD approach for different numbers of gates in Fig.~\ref{fig:digital_CD}. When the number of gates large, the digital CD protocol has higher fidelity than the standard Trotter protocol. The final fidelity is determined by the schedule $\lambda(t)$. Moreover, we see that even with fewer gates, the first-order CD approach has higher fidelity than the standard Trotter method.

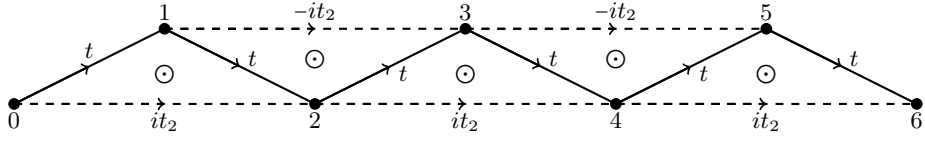
\begin{figure*}
\centering
\begin{tikzpicture}[scale=1]
\draw[thick] (0,0) -- (2,1);
\draw[thick] (2,1) -- (4,0);
\draw[thick] (4,0) -- (6,1);
\draw[thick] (6,1) -- (8,0);
\draw[thick] (8,0) -- (10,1);
\draw[thick] (10,1) -- (12,0);
\draw[->][thick] (0,0) -- (1,0.5);
\draw[->][thick] (2,1) -- (3,0.5);
\draw[->][thick] (4,0) -- (5,0.5);
\draw[->][thick] (6,1) -- (7,0.5);
\draw[->][thick] (8,0) -- (9,0.5);
\draw[->][thick] (10,1) -- (11,0.5);
\draw[dashed,->][thick] (0,0) -- (2,0);
\draw[dashed,->][thick] (4,0) -- (6,0);
\draw[dashed,->][thick] (8,0) -- (10,0);
\draw[dashed,->][thick] (2,1) -- (4,1);
\draw[dashed,->][thick] (6,1) -- (8,1);
\draw[dashed,thick] (0,0) -- (12,0);
\draw[dashed,thick] (2,1) -- (10,1);

\filldraw [black] (0,0) circle (2pt) node[anchor=north] {$0$};
\filldraw [black] (2,1) circle (2pt) node[anchor=south] {$1$};
\filldraw [black] (4,0) circle (2pt) node[anchor=north] {$2$};
\filldraw [black] (6,1) circle (2pt) node[anchor=south] {$3$};
\filldraw [black] (8,0) circle (2pt) node[anchor=north] {$4$};
\filldraw [black] (10,1) circle (2pt) node[anchor=south] {$5$};
\filldraw [black] (12,0) circle (2pt) node[anchor=north] {$6$};
\node[above] at (1,0.5) {$\tt$};
\node[anchor=west] at (3,0.6) {$\tt$};
\node[anchor=west] at (5,0.4) {$\tt$};
\node[anchor=west] at (7,0.6) {$\tt$};
\node[anchor=west] at (9,0.4) {$\tt$};
\node[anchor=west] at (11,0.6) {$\tt$};
\node[below] at (2,0) {$i t_2$};
\node[below] at (6,0) {$i t_2$};
\node[below] at (10,0) {$i t_2$};
\node[above] at (4,1) {$-i t_2$};
\node[above] at (8,1) {$-i t_2$};
\node at (2,0.4) {$\bigodot$};
\node at (4,0.6) {$\bigodot$};
\node at (6,0.4) {$\bigodot$};
\node at (8,0.6) {$\bigodot$};
\node at (10,0.4) {$\bigodot$};
\end{tikzpicture}
\caption{A 1d fermion chain with both nearest-neighbor (NN) and next-nearest-neighbor (NNN) hopping. The NN hopping terms are all $\tt$ while NNN terms are $i t_2$ and $-i t_2$ alternatively ($\tt$, $t_2$ are real). This corresponds to $\frac{\pi}{2}$ flux insertion in each triangle.  
}
\label{fig: 1d fermion chain}
\end{figure*}

\subsection{1d fermion chain with next-nearest-neighbor hopping terms}\label{sec: 1d fermion chain}

In this section, we discuss how to generate the time evolution of a 1d fermion chain with nearest-neighbor (NN) and next-nearest-neighbor (NNN) hopping terms using only two-site gates acting on neighboring sites on a fermionic digital quantum simulator.\footnote{The following derivation uses fermionic operators $c,c^\dagger$, which can be transformed to Pauli operators by the Jordan–Wigner transformation. Thus, this model can be realized by quantum computers with qubits using only two-qubit interactions.}

Consider a 1d fermion chain with one complex fermion on each site $j$. The fermion operators $c_j$, $c^\dagger_j$ satisfy the canonical anticommutation relations
\begin{equation}
    \{c_j, c^\dagger_k \} = \delta_{jk},\quad \{c_j, c_k \} = \{c^\dagger_j, c^\dagger_k \}=0.
\end{equation}
We partition the hopping terms into two sets: those between sites $2j$, $2j+1$ and between sites $2j+1$, $2j+2$. More explicitly, we define
\begin{equation}
    \begin{split}
        H_{0} &:= \big(c^\dagger_0 c_1 + c^\dagger_2 c_3 + \cdots + c^\dagger_{\nq -2} c_{\nq -1} \big) + \hc \\
        H_{1} &:= \big(c^\dagger_1 c_2 + c^\dagger_3 c_4 + \cdots + c^\dagger_{\nq -1} c_{0}\big) + \hc,
    \end{split}
\end{equation}
where we use periodic boundary conditions and assume that the total number of sites $\nq$ is even.  Notice that the terms in $H_0$ pairwise commute, so the time evolution of $H_0$ is exactly the product of the time evolutions of the individual terms, and similarly for $H_1$. By choosing $A = i H_0$, $B = i H_1$, $\alpha = -\tt T$, and $\beta = -t_2 T$ in Eqs.~\eqref{eq:desired_gate} and \eqref{eq: product formula}, we have
\begin{equation}
    f_{\frac{\beta n}{\alpha^2}}\left(\frac{\alpha}{n}\right) = \exp\left(- i H_{\text{eff}} \, \frac{T}{n} + O\left(\frac{LT^2}{n^2}\right)\right)
\label{eq: Trotter steps of H_eff}
\end{equation}
where
\begin{align}\label{eq:eff_commut}
        H_{\text{eff}} &= \tt (H_0 + H_1) + i t_2 [H_0, H_1] \nonumber\\
        &= \tt \big(c^\dagger_0 c_1 + c^\dagger_1 c_2 + c^\dagger_2 c_3 + c^\dagger_3 c_4 + \cdots + \hc\big)\\[3pt]
        &\quad+ t_2\big(i c^\dagger_0 c_2 - i c^\dagger_1 c_3 + i c^\dagger_2 c_4 - i c^\dagger_3 c_5 + \cdots + \hc \big). \nonumber
\end{align}
Repeating the product formula $n$ times, we find
\begin{equation}
    f_{\frac{\beta n}{\alpha^2}}\left(\frac{\alpha}{n}\right)^n = \exp\left( -i H_{\text{eff}} T + O\left(\frac{LT^2}{n}\right)\right)
\label{eq: simulation of Heff}
\end{equation}

The effective Hamiltonian $H_{\text{eff}}$ is shown in Fig.~\ref{fig: 1d fermion chain}. It contains NN hopping terms with amplitude $\tt$ and NNN hoppings terms with amplitude $t_2$ and alternating $i$ and $-i$ factors. Physically, this Hamiltonian corresponds to the insertion of a $\frac{\pi}{2}$ flux in each triangle. If we transform the NNN term $c_j^\dagger c_{j+2}$ to a qubit representation, it corresponds to a 3-qubit interaction $\sigma_j^+\sigma_{j+1}^z\sigma_{j+2}^-$ as occurs in lattice gauge theories \cite{Floquet_LGT}.\footnote{In certain systems, there may be other ways of efficiently generating such a 3-qubit term. For example, in the Google quantum computer, this term can be realized using a few single-qubit and iSwap gates \cite{Google_1d_fermion2020}. In general, the best circuit depends on the details of the physical platform. The preceding discussion provides an alternative simulation method, which may benefit some quantum simulators.}

To simulate $H_0$ or $H_1$, we use $\frac{L}{2}$ gates. To simulate the time evolution of $H_{\text{eff}}$, we use $6n \times \frac{L}{2} = 3 n L$ gates, which is proportional to the number of steps $n$ and the chain length $L$. 
Thus, using only nearest-neighbor terms (or 2-qubit gates in the qubit representation), we are able to simulate the next-nearest-neighbor terms (or 3-qubit gates) with the same number of gates as in standard Trotterization of a Hamiltonian with only nearest-neighbor terms.\footnote{In the standard Trotter approach, the evolution time for each step is the same and the effective Hamiltonian is the sum of the simulated terms. In the algorithm discussed here, the evolution times are fine-tuned such that the next-nearest-neighbor terms appear in the effective Hamiltonian.} We do not require any gate decomposition of 3-qubit interactions.

\subsection{Fractional quantum Hall phases on lattices}\label{sec: Kapit-Mueller}

In Ref.~\cite{KM_model}, Kapit-Mueller showed that the addition of an appropriate next-nearest-neighbor hopping terms to magnetic models on lattices can significantly flatten the lowest band. Such a band flattening can further stabilize and increase the gap of lattice quantum Hall states such as Laughlin states. In this section, we discuss how to simulate the Kapit-Mueller Hamiltonian with nearest-neighbor (NN) and next-nearest-neighbor (NNN) hopping terms on a two-dimensional square lattice using only the time evolution operators of nearest-neighbor hopping terms.

Using our product formula for both sum and commutator, as in Eq.~\eqref{eq: product formula}, we can simulate the Hamiltonian of the Kapit-Mueller model. The general form of the Kapit-Mueller Hamiltonian can be regarded as a variation of the Hofstadter Hamiltonian \cite{Hofstadter_model}, which involves not only NN hopping but also long-range hopping. The long-range hopping ensures a flat band, which can be regarded as a degenerate Landau levels to stabilize fractional quantum Hall states. By truncating to only the NN and NNN terms, the Kapit-Mueller model simplifies the interactions while still demonstrating features of fractional quantum Hall phases.

\begin{figure}
	\centering
	\begin{tikzpicture}[scale=1.5]
	\draw[thick,red] (0,0)--(1,0);
	\draw[thick,red] (1,1)--(2,1);
	\draw[thick,red] (0,2)--(1,2);
	
	\draw[thick,blue] (1,0)--(2,0);
	\draw[thick,blue] (0,1)--(1,1);
	\draw[thick,blue] (1,2)--(2,2);
	
	\draw[thick,green] (1,0)--(1,1);
	\draw[thick,green] (0,1)--(0,2);
	\draw[thick,green] (2,1)--(2,2);
	
	\draw[thick,cyan] (0,0)--(0,1);
	\draw[thick,cyan] (2,0)--(2,1);
	\draw[thick,cyan] (1,1)--(1,2);
	
	\filldraw[black] (0,0) circle (2pt);
	\filldraw[black] (1,0) circle (2pt);
	\filldraw[black] (2,0) circle (2pt);
	\filldraw[black] (0,1) circle (2pt);
	\filldraw[black] (0,2) circle (2pt);
	\filldraw[black] (1,1) circle (2pt);
	\filldraw[black] (1,2) circle (2pt);
	\filldraw[black] (2,1) circle (2pt);
	\filldraw[black] (2,2) circle (2pt);
	
	\draw[thick, ->] (0.8,0.5) arc (0:270:0.3cm) node[above=.1cm] {\Large $\phi$};
	\draw[thick, ->] (0.8,1.5) arc (0:270:0.3cm) node[above=.1cm] {\Large$\phi$};
	\draw[thick, ->] (1.8,0.5) arc (0:270:0.3cm) node[above=.1cm] {\Large$\phi$};
	\draw[thick, ->] (1.8,1.5) arc (0:270:0.3cm) node[above=.1cm] {\Large$\phi$};

	\end{tikzpicture}
	\caption{\label{KM_figure} (Color online) Square lattice in a uniform magnetic field. The red links belong to $H_1$, the blue links belong to $H_2$, the green links belong to $H_3$, and the light blue links belong to $H_4$.}
\end{figure}
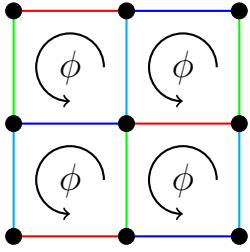

Specifically, the Kapit-Mueller Hamiltonian is of the form
\begin{eqs}\label{allequations}
H_{\text{KM}}&=\sum_{\langle j, k\rangle,~\llangle j,k\rrangle}J(z_j,z_k)a_j^\dagger a_k, ~ \text{where} \\
J(z_j,z_k)&=K(z)e^{(\pi/2)(z_jz^*-z_j^*z)\phi}\\
K(z)&=t\times (-1)^{x+y+xy}e^{-\frac{\pi}{2}[(1-\phi)|z|^2]}
\end{eqs}
with $z_j=x_j+iy_j$ denoting the position of the $j$th site, $z=z_k-z_j$ representing the displacement from the $j$th site to the $k$th site, and $\langle j,k \rangle$ indicating that $j,k$ are NN sites, and $\llangle j,k \rrangle$ indicating that $j,k$ are NNN sites. The $a_j^\dagger$ operator is a bosonic creation operator acting on site $j$, and $\phi$ is the magnetic flux inside each plaquette, as shown in Fig.~\ref{KM_figure}. The hopping phase factor $(z_jz^*-z_j^*z)\phi$ corresponds to a vector potential in the symmetric gauge.

For convenience, we use Cartesian coordinates $(m,n)$ to label lattice sites and rescale $\pi\phi$ to $\phi$. The lattice constant is set to 1.
We divide all the nearest-neighbor hopping terms into four parts:
\begin{eqs}\label{3B:nearest_neighbor}
    H_1&=-J\sum_{m+n \text{ even}}e^{-in\phi}a_{m+1,n}^\dagger a_{m,n}+\hc, \\
    H_2&=-J\sum_{m+n \text{ odd}}e^{-in\phi}a_{m+1,n}^\dagger a_{m,n}+\hc, \\
    H_3&=-J\sum_{m+n \text{ odd}}a_{m,n+1}^\dagger a_{m,n}+\hc,\\
    H_4&=-J\sum_{m+n \text{ even}}a_{m,n+1}^\dagger a_{m,n}+\hc
\end{eqs}

Using the bosonic commutation relation $[a_i,a_j^\dagger]=\delta_{i,j}$, we see that the commutator between two hopping terms that overlap on site $j$ is
\begin{equation}
    -i[K_1a_i^\dagger a_j+\hc,K_2a_j^\dagger a_k+\hc]=-iK_1K_2a_i^\dagger a_k+\hc
    \label{3B:commutator}
\end{equation}
Thus commutators between NN hopping terms can generate NNN hopping. We compute the following commutators between different NN hopping terms:
\begin{widetext}
    \begin{eqs}\label{3B:NNN_hopping}
    &-i[H_1, H_3]  =- J^2 \sum_{m+n=\text{odd}}ie^{-i(n+1)\phi}a_{m+1,n+1}^\dagger a_{m,n}+\hc +J^2\sum_{m+n=\text{even}}ie^{-in\phi}a_{m+1,n+1}^\dagger a_{m,n}+\hc,\\
    &-i[H_1,H_4]  =-J^2\sum_{m+n=\text{even}}ie^{-in\phi}a_{m+1,n}^\dagger a_{m,n+1}+\hc+J^2\sum_{m+n=\text{odd}}ie^{-i(n+1)\phi}a_{m+1,n}^\dagger a_{m,n+1}+\hc,\\
    &-i[H_2,H_3] =-J^2\sum_{m+n=\text{odd}}ie^{-in\phi}a_{m+1,n}^\dagger a_{m,n+1}+\hc +J^2\sum_{m+n=\text{even}}ie^{-i(n+1)\phi}a_{m+1,n}^\dagger a_{m,n+1}+\hc, \\
    &-i[H_2,H_4] =-J^2\sum_{m+n=\text{even}}ie^{-i(n+1)\phi}a_{m+1,n+1}^\dagger a_{m,n}+\hc +J^2\sum_{m+n=\text{odd}}ie^{-in\phi}a_{m+1,n+1}^\dagger a_{m,n}+\hc
    \end{eqs}
Therefore,
\begin{eqs}
    &-i[H_1,H_3]-i[H_2,H_4]+i[H_1,H_4]+i[H_2,H_3]\\
    =&-2J^2\sin\left(\frac{\phi}{2}\right)\sum_{m,n}e^{-i(n+\frac{1}{2})\phi}\left(a_{m+1,n+1}^\dagger a_{m,n}  + a_{m+1,n}^\dagger a_{m,n+1} \right)+\hc\\
    =&-i[H_1-H_2,H_3-H_4].
\end{eqs}
\end{widetext}

Comparing the phase factors of NN hopping in Eq.~\eqref{3B:nearest_neighbor} and NNN hopping in Eq.~\eqref{3B:NNN_hopping}, we find the effective Hamiltonian in terms of $H_1$, $H_2$, $H_3$, $H_4$ and their commutators,
\begin{equation}
    H_{\text{eff}}=
    H_1+H_2+H_3+H_4-iJ^\prime[H_1-H_2,H_3-H_4]
\end{equation}
which describes a system with NN and NNN hopping where the magnetic flux inside each plaquette is $\phi/2$. Our product formula naturally generates the phase factor corresponding to a uniform magnetic field applied to the lattice. Since we can control the coefficient of the commutator term in the product formula, $H_{\text{eff}}$ is equivalent to Kapit-Mueller Hamiltonian
with the choice $J^\prime=\frac{\exp(\phi/4-\pi/2)}{2J\sin(\phi/2)}$, which is fine tuned to realize a flat band.
We can use our product formula Eq.~\eqref{eq: modified product formula} to simulate $H_{\text{eff}}$ by setting $A=i(H_1-H_2)$, $B=i(H_3-H_4)$, $C=i(2H_2+2H_4)$, and taking $\alpha=1$, $\beta=J'$.

\section{Discussion} \label{sec: discussions}

We conclude by discussing some possible future research directions.

    \paragraph*{Trajectories of product formulas:}
    Given a product formula
    \begin{eqs}
        e^{ p_1 x A} e^{p_2 x B} e^{ p_3 x A} e^{p_4 x B} e^{ p_5 x A} e^{p_6 x B} \ldots,
    \end{eqs}
    we can plot the ``time evolution trajectory'' of the coefficients of $A$ and $B$, namely $(p_1, p_1+ p_3, p_1+p_3+p_5, \dots)$ and $(p_2, p_2 + p_4, p_2 + p_4 + p_6, \dots)$, respectively. Ref.~\cite{NM05} studies product formulas for the sum $A+B$, with the time evolution trajectory starting from $t=0$ and ending at $t=1$. It suggests that a good formula product formula for the sum should have the entire time evolution trajectory inside the ``allowed'' interval $[0,1]$, since times outside this interval do not correspond to the evolution under consideration.

\begin{figure}
        \centering
        \includegraphics[width=0.45\textwidth]{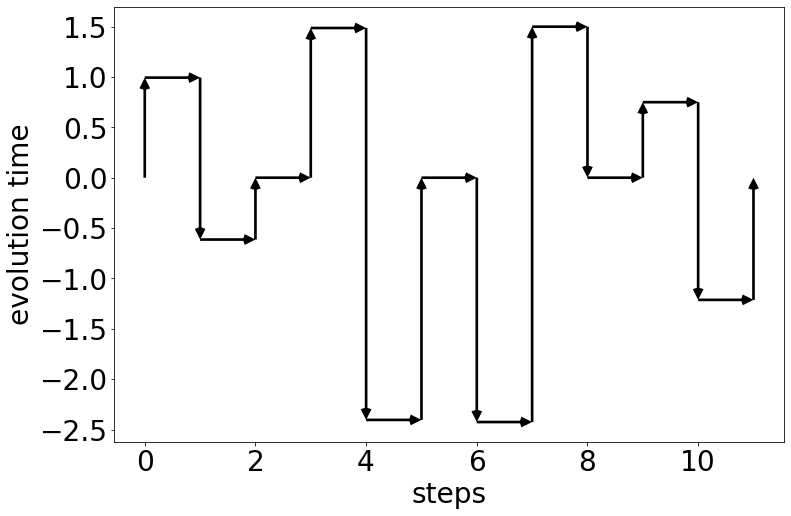}\\[1ex]
        \includegraphics[width=0.45\textwidth]{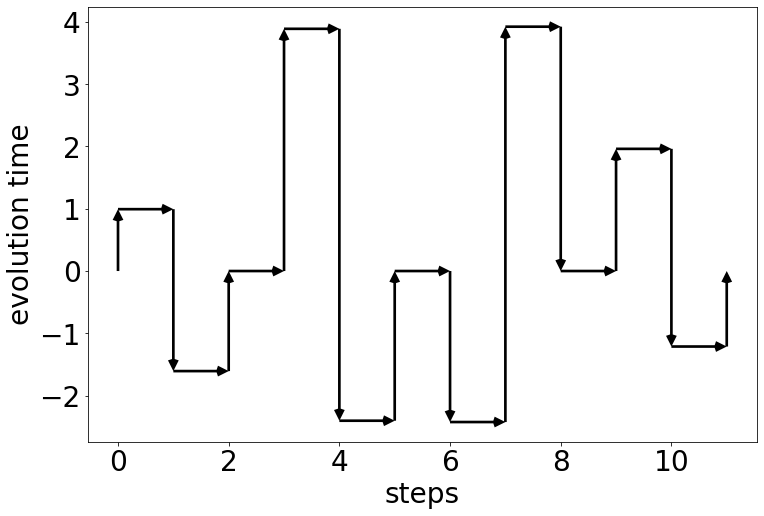}
        \includegraphics[width=0.45\textwidth]{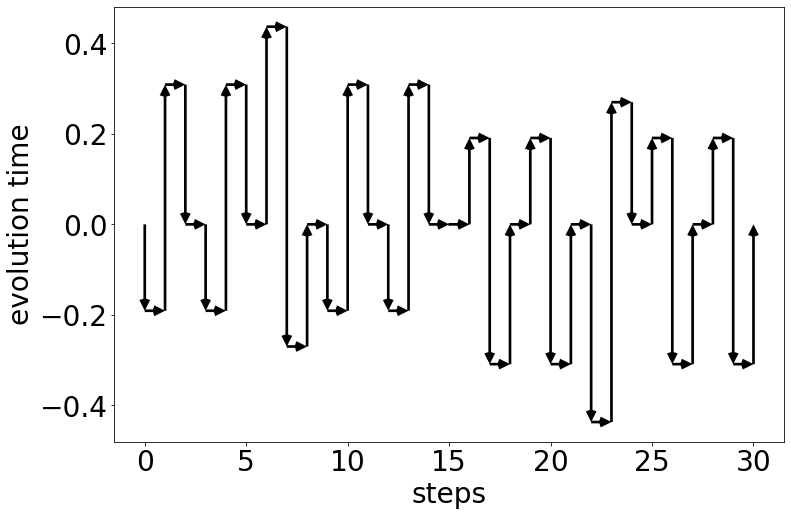}\\[1ex]
        \includegraphics[width=0.45\textwidth]{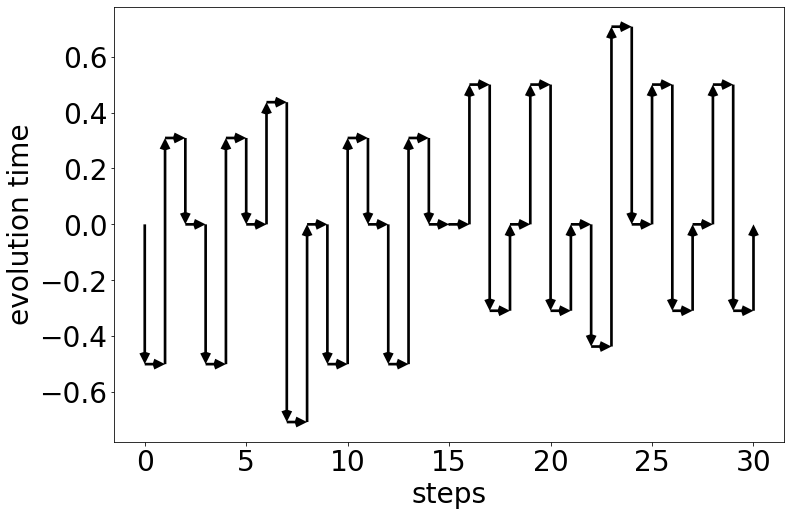}
        \caption{Time evolution trajectories of product formulas. From top to bottom, $Q_5(1)$ ($\sqrt{4}$-copy) for $A$ and $B$ and $\mathcal{G}_5(1)$ ($\sqrt{10}$-copy) for $A$ and $B$.}
        \label{fig:trajectories}
    \end{figure}

   For commutators, the time evolution trajectory starts and ends at $t=0$, so we do not have an ``allowed'' interval.
   However, we can still plot the time evolution trajectories, as shown in Fig.~\ref{fig:trajectories} for the $\sqrt{4}$-copy and $\sqrt{10}$-copy product formulas.  We see that $\sqrt{10}$-copy product formula has a smaller range for the time evolution trajectories of $A$ and $B$. This may explain why the $\sqrt{10}$-copy product formula performs better than the $\sqrt{4}$-copy product formula. Similar considerations hold for other formulas (for example, the ranges of the trajectories for the $\sqrt{5}$-copy formula are intermediate between those of the $\sqrt{4}$- and $\sqrt{10}$-copy formulas).
   It might be useful to develop a more general and quantitative understanding of how the trajectories of a product formula affect its performance.
    
    \paragraph*{Optimal recursive relation:} We argue that the $\sqrt{4}$-copy recursive relation may use the smallest possible number of copies to generate higher-order product formulas.  In other words, it seems unlikely that a recursive relation could use fewer than $2$ copies of an $n$th-order product formula to generate an $(n+1)$st-order product formula. To obtain a $p$th-order product formula, the coefficients of $p$th-order commutators must be eliminated. If $p$ is a prime, the number of independent commutators is $\frac{2^p-2}{p}$, which is the dimension of the graded component of a free Lie algebra with length\footnote{The length is the number of times $A$ and $B$ occur in Lie brackets. For example, there are two terms of length $3$: $[A,[A,B]]$ and $[B,[B,A]]$.} $p$ \cite{reutenauer93, bourbaki08}.
    A $(2-\eps)$-copy recursive formula would use a number of exponentials proportional to $(2-\eps)^p \ll \frac{2^p-2}{p}$, which would mean using far fewer parameters than the number of polynomial equations to be solved.
    Thus we conjecture that there is no $(2-\eps)$-copy recursive formula.
    
    \paragraph*{Direct derivation of a $4$th-order commutator product formula:} To better understand possible direct constructions of commutator product formulas, we would like to solve the polynomial equations for the $4$th-order case, namely\footnote{While the number of terms in the product formula is finite, we do not indicate where it ends since we do not know a priori how many terms are required.}
    \begin{eqs}
        &e^{ p_1 x A} e^{p_2 x B} e^{ p_3 x A} e^{p_4 x B} e^{ p_5 x A} e^{p_6 x B} \ldots \\
        &= e^{x^2 [A,B]} + O(x^5).
    \end{eqs}
    
    We first revisit the polynomial equations Eq.~\eqref{eq: polynomial equation for the 3rd order commutators} for the 3rd-order product formula. We can rewrite these equations as
    \begin{equation}
    \begin{aligned}
        \A &= 0, &
        \B &= 0, &
        \B \A &= -1,  \\
        \A \B \A &= 0,  &
        \B \A \B &= 0,  \\
        \end{aligned}
    \end{equation}
    where
    \begin{eqs}
        \A := \sum_{i \text{~odd}} p_i = p_1 + p_3 + p_5 +\cdots
    \end{eqs}
    represents the sum of all coefficients of the $A$ term
    and
    \begin{eqs}
        \B := \sum_{i \text{~even}} p_i = p_2 + p_4 + p_6 +\cdots
    \end{eqs}
    represents the sum of all coefficients of the $B$ term.
    Then 
    \begin{eqs}
        \B \A := \sum_{i \text{~even}, \, j \text{~odd}, \, i<j} p_i p_j.
    \end{eqs}
    is the sum of all coefficients of the $BA$ term, and similarly
    \begin{eqs}
        \A \B \A &:= \sum_{i \text{~odd},\, j \text{~even},\, k \text{~odd},\, i<j<k} p_i p_j p_k, \\
        \B \A \B &:= \sum_{i \text{~even},\, j \text{~odd},\, k \text{~even},\, i<j<k} p_i p_j p_k.
    \end{eqs}
    
    Following the same strategy, we can derive polynomial equations for the $4$th-order commutator product formula:
    \begin{equation}
    \begin{aligned}
        \A &= 0, &
        \B &= 0, \\
        \B \A &= -1, &  
        \A \B \A &= 0,  \\
        \B \A \B &= 0, & 
        \A^2 \B \A &= 0, \\
        \B^2 \A \B &= 0, &
        \A \B \A \B - \B \A \B \A &= 0,
        \end{aligned}
    \end{equation}
    where we have defined
    \begin{eqs}
        \A^2 \B \A &:= \sum_{i \text{~odd},\, j \text{~even},\, k \text{~odd},\, i<j<k} \frac{1}{2 }p_i^2 p_j p_k \\
        &\quad+\sum_{\substack{i \text{~odd},\, j \text{~odd},\, k \text{~even},\\ l \text{~odd},\, i<j<k<l}} p_i p_j p_k p_l, \\
        \B^2 \A \B &:= \sum_{i \text{~even},\, j \text{~odd},\, k \text{~even},\, i<j<k} \frac{1}{2 }p_i^2 p_j p_k \\
       &\quad +\sum_{\substack{i \text{~even},\, j \text{~even},\, k \text{~odd}, \\ l \text{~even},\, i<j<k<l}} p_i p_j p_k p_l, \\
        \A \B \A \B &:= \sum_{\substack{i \text{~odd},\, j \text{~even},\, k \text{~odd}, \\ l \text{~even},\, i<j<k<l}} p_i p_j p_k p_l, \\
        \B \A \B \A &:= \sum_{\substack{i \text{~even},\, j \text{~odd},\, k \text{~even},\\ l \text{~odd},\, i<j<k<l}} p_i p_j p_k p_l.
    \end{eqs}
    However, we do not have an analytical solution for these polynomial equations.
    
    \paragraph*{Applications of counterdiabatic driving:} In Sec.~\ref{sec: CD}, we gave a two-qubit example to demonstrate the potential effectiveness of counterdiabatic driving in digital quantum computers. We believe this approach can be used to prepare ground states of spin-chain systems with high fidelity. Looking beyond quantum simulation, there might exist other efficient quantum algorithms based on counterdiabatic driving.


\begin{acknowledgements}
We thank Mathias Van Regemortel, Bowen Yang, and Yixu Wang for helpful discussions.
Y.-A.C.\ is supported by the JQI fellowship.
A.M.C.\ received support from the National Science Foundation (grant CCF-1813814 and QLCI grant OMA-2120757) and the Department of Energy, Office of Science, Office of Advanced Scientific Computing Research, Accelerated Research in Quantum Computing program. Y.X., H.K., and M.H.\ were supported by ARO W911NF-15-1-0397, National Science Foundation QLCI grant OMA-2120757, AFOSR-MURI FA9550- 19-1-0399,  Department of Energy QSA program, and the Simons Foundation. 
\end{acknowledgements}

\appendix

\section{The operator differential method and commutator product formulas} \label{sec: operator differential and product formula}

In this section, we introduce the operator differential method and show how it can be used to derive product formulas. We have \cite{NM05}
\begin{equation}
    e^{p_1 x A} e^{p_2 x B} e^{p_3 x A} e^{p_4 x B} e^{p_5 x A} e^{p_6 x B} = e^{\Phi (x)}
\end{equation}
with
\begin{align}
        &\Phi (x) \nonumber\\
        &= \sum_{k=0}^\infty \frac{(-1)^k}{k+1} \int_0^x (e^{p_1 t \da} e^{p_2 t \db} e^{p_3 t \da}
        e^{p_4 t \db} e^{p_5 t \da} e^{p_6 t \db} -1 )^k \nonumber\\
        &\quad \times (p_1 A + e^{p_1 t \da} p_2 B + e^{p_1 t \da} e^{p_2 t \db} p_3 A \cdots)\, \mathrm{d}t,
\label{eq: general product formula}
\end{align}
where $\delta_A O := [A, O]$ is called the operator differential. For example, the $x$ term in $\Phi(x)$ comes from integrating the constant term in Eq.~\eqref{eq: general product formula}:
\begin{eqs}
    &\int_0^x (p_1 A + p_2 B + p_3 A + p_4 B + p_5 A + p_6 B) \mathrm{d}t \\
    & = (l A + m B) x,
\end{eqs}
where
\begin{eqs}
    l &:= p_1 + p_3 + p_5, \\
    m &:= p_2 + p_4 + p_6.
\end{eqs}
Notice that only the $k=0$ part contributes to the constant term in the integrand. The $x^2$ term in $\Phi(x)$ comes from the $t$ terms in the integrand, which have two contributions: from the $k=0$ part, we have
\begin{eqs}
    &\int_0^x \mathrm{d}t \,  t  \left[p_1(p_2+p_4+p_6) + p_3(p_4 + p_6) + p_5 p_6 \right] \delta_A B \\
    & \quad + t \left[p_2(p_3+p_5) + p_4 p_5 \right] \delta_B A  \\
    &= \frac{x^2}{2} \left[  p_1(p_2+p_4+p_6) + p_3(p_4 + p_6) + p_5 p_6  \right. \\
    & \quad \left. - p_2(p_3+p_5) - p_4 p_5 \right] \da B ,
\end{eqs}
and from the $k=1$ part, we have
\begin{eqs}
    & - \frac{1}{2}\int_0^x \mathrm{d}t \, t (p_2 + p_4 + p_6) (p_1 + p_3 + p_5) \db A \\
    & \quad \quad + t (p_1 + p_3 + p_5) (p_2 + p_4 + p_6) \da B \\
    & = 0.
\end{eqs}
The $x^2$ term is therefore
\begin{eqs}
    \frac{x^2}{2} (l m - 2q) \da B,
\label{eq: x^2 term}
\end{eqs}
where
\begin{eqs}
    q := p_2 p_3 + p_2 p_5 + p_4 p_5.
\end{eqs}
The $x^3$ term in $\Phi(x)$ has three contributions. The first comes from the $k=0$ part:
\begin{eqs}
    &\int_0^x \mathrm{d}t \frac{t^2}{2} [ p_1^2 (p_2 + p_4 + p_6) + p_3^2 (p_4 + p_6) + p_5^2 p_6  \\
    & \quad \quad \quad \quad  + 2p_1 p_3 (p_4+p_6) +2 (p_1 + p_3) p_5 p_6 \\
    & \quad \quad \quad \quad - 2 p_1 p_2 (p_3 + p_5) - 2 (p_1 + p_3) p_4 p_5] \delta_A^2 B \\
    & \quad \quad + \frac{t^2}{2} [p_2^2 (p_3 + p_5) + p_4^2 p_5 + 2 p_2 p_4 p_5 \\
    & \quad \quad \quad \quad -2 p_2 p_3 (p_4+p_6) - 2(p_2 + p_4) p_5 p_6] \db^2 A. \\
\end{eqs}
Defining
\begin{eqs}
    r &:= p_1 p_2 p_3 + p_1 p_2 p_5 + p_1 p_4 p_5 + p_3 p_4 p_5\\
    s &:= p_2 p_3 p_4 + p_2 p_3 p_6 + p_2 p_5 p_6 + p_4 p_5 p_6,
\end{eqs}
the $k=0$ part can be simplified as
\begin{eqs}
    &\int_0^x \mathrm{d}t \frac{t^2}{2} [ (l (lm-q) -3r) \da^2 B + (mq - 3s) \db^2 A ] \\
    &= \frac{x^3}{6} [ (l (lm-q) -3r) \da^2 B + (mq - 3s) \db^2 A ].
\end{eqs}
Similarly, for $k=1$ we have
\begin{align}
    &-\frac{1}{2}\int_0^x \mathrm{d}t \, t (l \da + m \db) t[ (lm-2q)\da B ]  \nonumber\\
    & \quad \quad + \frac{1}{2} t^2 (l^2 \da^2 + m^2 \db^2 + 2(lm-q) \da \db + 2q \db \da) \nonumber\\
    & \quad \quad ~~ \cdot (lA + mB) \nonumber\\
    & = -\frac{x^3}{6} \left[ \left(\frac{1}{2} l^2 m - lq\right) \da^2 B - \left(\frac{1}{2} m^2 l - mq\right) \db^2 A \right]
\end{align}
and for $k=2$ we have
\begin{eqs}
    & \frac{1}{3} \int_0^x \mathrm{d}t \, t^2 (l \da + m \db)^2 (l A + m B) \\
    &= \frac{1}{3} \int_0^x t^2 \delta_{l A + m B}^2 (l A + m B) = 0.
\end{eqs}
Overall, the $x^3$ term is
\begin{eqs}
    \frac{x^3}{6} \left[ \left(\frac{l^2 m}{2} -3r\right) \da^2 B +\left(\frac{m^2 l}{2} -3s\right) \db^2 A \right].
\label{eq: x^3 term}
\end{eqs}

\subsection{Pure commutators}

To give a pure commutator formula, we would like to find $(p_1, p_2, p_3, p_4, p_5, p_6)$ such that
\begin{eqs}
    \Phi(x) = R [A,B] x^2 + O(x^4)
\end{eqs}
for some constant $R$. For the first-order term to vanish, we require
\begin{equation}
    \begin{split}
        l = p_1 + p_3 + p_5 &= 0 \\ 
        m = p_2 + p_4 + p_6 &= 0.
    \end{split}
\end{equation}
The $x^2$ term, Eq.~\eqref{eq: x^2 term}, and $x^3$ term, Eq.~\eqref{eq: x^3 term}, contribute
\begin{eqs}
    -x^2 q \da B - \frac{x^3}{2} ( r \da^2 B + s \db^2 A).
\end{eqs}
To eliminate the third-order term, we need to solve
\begin{equation}
    \begin{split}
        l &= p_1 + p_3 + p_5 = 0, \\ 
        m &= p_2 + p_4 + p_6 = 0, \\
        q &= p_2 p_3 + p_2 p_5 + p_4 p_5 = -1, \\
        r &= p_1 p_2 p_3 + p_1 p_2 p_5 + p_1 p_4 p_5 + p_3 p_4 p_5 = 0, \\
        s &= p_2 p_3 p_4 + p_2 p_3 p_6 + p_2 p_5 p_6 + p_4 p_5 p_6 = 0.
    \end{split}
\end{equation}
One can check that the following choice satisfies the equations:
\begin{equation}
\begin{aligned}
    p_1 &=\frac{\sqrt{5}-1}{2}, & p_2 &=\frac{\sqrt{5}-1}{2}, & p_3&=-1, \\
    p_4 &= -\frac{\sqrt{5}+1}{2}, & p_5&=\frac{3-\sqrt{5}}{2}, & p_6&= 1.
\end{aligned}
\end{equation}
Thus we have the explicit product formula Eq.~\eqref{eq: 6 exp product formula for commutator}.

\subsection{Sums and commutators}

Now consider a case where we include both linear and commutator terms, namely
\begin{eqs}
    \Phi (x) = (A+B) x + R[A,B] x^2 + O(x^4),
\end{eqs}
for an arbitrary constant $R$. The first-order term requires
\begin{eqs}
    l &= p_1 + p_3 + p_5 = 1, \\ 
    m &= p_2 + p_4 + p_6 = 1, \\
\end{eqs}
and the $x^2$ and $x^3$ terms are
\begin{eqs}
    \frac{x^2}{2}(1-2q) + \frac{x^3}{6} \left[ \left(\frac{1}{2} -3r\right) \da^2 B + \left(\frac{1}{2} - 3s\right) \db^2 A \right],
\end{eqs}
which agrees with the derivation in Ref.~\cite{NM05}. Therefore, the equations to be solved are
\begin{align}
    l &= p_1 + p_3 + p_5 = 1, \nonumber\\ 
    m &= p_2 + p_4 + p_6 = 1, \nonumber\\
    q &= p_2 p_3 + p_2 p_5+ p_4 p_5 = -R +\frac{1}{2}, \nonumber\\
    r &= p_1 p_2 p_3 + p_1 p_2 p_5 + p_1 p_4 p_5 + p_3 p_4 p_5 = \frac{1}{6}, \nonumber\\
    s &= p_2 p_3 p_4 + p_2 p_3 p_6 + p_2 p_5 p_6 + p_4 p_5 p_6 = \frac{1}{6}.
\end{align}

\section{Existence of the \texorpdfstring{$\sqrt{4}$}{}-copy recursive formula} \label{sec: existence of solution}

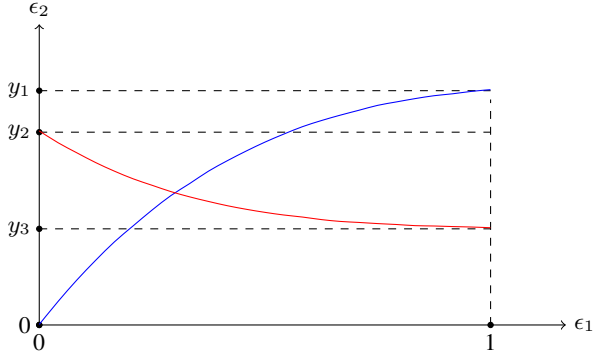
\begin{figure}
    \centering
    \begin{tikzpicture}[scale=2]
    \draw[->] (0,0) -- (3.5,0) node[right] {$\epsilon_1$};
    \draw[->] (0,0) -- (0,2) node[above] {$\epsilon_2$};
    \filldraw[black] (0,0) circle (0.5 pt) node[anchor=north] {0};
    \filldraw[black] (0,0) circle (0.5 pt) node[anchor=east] {0};
    \filldraw[black] (3,0) circle (0.5 pt) node[anchor=north] {1};
    \filldraw[black] (0,1.2824757) circle (0.5 pt) node[anchor=east] {$y_2$};
    \filldraw[black] (0,1.5585595) circle (0.5 pt) node[anchor=east] {$y_1$};
    \filldraw[black] (0,0.6395278) circle (0.5 pt) node[anchor=east] {$y_3$};
    \draw[dashed] (0,1.2824757) -- (3,1.2824757);
    \draw[dashed] (0,1.5585595) -- (3,1.5585595);
    \draw[dashed] (0,0.6395278) -- (3,0.6395278);
    \draw[dashed] (3,0) -- (3,1.5);
    \draw[scale=5, domain=0:0.6, smooth, variable=\x, blue] plot ({\x}, {10*(2-(15+(1-\x)^4)^(1/4))});
    \draw[scale=5, domain=0:0.6, smooth, variable=\x, red] plot ({\x}, {10*(2-(31+(-1+\x)^5)^(1/5))});
    \end{tikzpicture}
\caption{The blue curve is the solution of Eq.~\eqref{eq: first cd equation}, a monotonically increasing function between $(0,0)$ and ($1, y_1$). The red curve is the solution of \eqref{eq: second cd equation}, a monotonically decreasing function between $(0,y_2)$ and $(1,y_3)$. They intersect in the region $0 < \eps_1 < 1$. The plot shows the case $k=2$, but the curves are qualitatively similar for any $k$.}
\label{fig: e1-e2-solution}
\end{figure}

In this section, we prove that a real solution of
\begin{eqs}
    a^{n+1}-b^{n+1}+c^{n+1}-d^{n+1} &= 0 \\
    a^{n+2}-b^{n+2}+c^{n+2}-d^{n+2} &= 0 \\
    a^2-b^2+c^2-d^2 & \neq 0
\end{eqs}
exists for any odd $n=2k-1$. We first take $a=1$ and $b=2$, giving
\begin{align}
    c^{2k}-d^{2k} &= 2^{2k}-1, \label{eq: first cd equation} \\ 
    c^{2k+1}-d^{2k+1} &= 2^{2k+1} - 1. \label{eq: second cd equation}
\end{align}
The solution $(c,d)=(2,1)$ is trivial since $a^2-b^2+c^2-d^2 = 0$. We show that there exists a solution of the form $(c,d) = (2- \eps_2, -1+\eps_1)$ with $ 1 \geq \eps_1, \eps_2 \geq 0$.

For any positive integer $k$, the solutions of Eqs.~\eqref{eq: first cd equation} and \eqref{eq: second cd equation} line on curves in the $(\eps_1,\eps_2)$ plane.
We can check that the solutions of Eq.~\eqref{eq: first cd equation} include the two points $(\eps_1, \eps_2) = (0, 0),\,(1, y_1)$ with $y_1 := 2-(2^{2k}-1)^{\frac{1}{2k}}$. Similarly, the solutions of Eq.~\eqref{eq: second cd equation} include $(\eps_1, \eps_2) = (0, y_2)$ and $(1, y_3)$ with $y_2 = 2- (2^{2k+1}-2)^{\frac{1}{2k+1}}$ and $y_3 = 2- (2^{2k+1}-1)^{\frac{1}{2k+1}}$. Notice that $y_1 > y_2 > y_3$. As shown in Fig.~\ref{fig: e1-e2-solution}, the curve for Eq.~\eqref{eq: first cd equation} is monotonically increasing and the curve for Eq.~\eqref{eq: second cd equation} is monotonically decreasing, so these curves must intersect in the region $0<\eps_1<1$, providing a simultaneous solution of the two equations.

The final step is to show $a^2-b^2+c^2-d^2 \neq 0$, or equivalently,
\begin{eqs}
    (2-\eps_2)^2 - (1-\eps_1)^2 \neq 3.
\end{eqs}
It suffices to show that
\begin{eqs}
    (2-\eps_2)^2 - (1-\eps_1)^2 = 3
\label{eq: curve 1}
\end{eqs}
and 
\begin{eqs}
    (2-\eps_2)^{2k} - (1-\eps_1)^{2k}=2^{2k}-1
\label{eq: curve 2}
\end{eqs}
have no intersection in the interval $ 0<\eps_1 < 1$ for $k>1$. For each equation, we can think of $\eps_2$ as a function of $\eps_1$, i.e.,
\begin{eqs}
    f_1 (x) &:= 2 - [3+(1-x)^2]^{\frac{1}{2}},\\
    f_2 (x) &:= 2 - [2^{2k}-1+(1-x)^{2k}]^{\frac{1}{2k}},
\end{eqs}
corresponding to Eq.~\eqref{eq: curve 1} and Eq.~\eqref{eq: curve 2}, respectively.
Now consider their derivatives:
\begin{eqs}
    f'_1 (x) &= \frac{1-x}{[ 3 + (1-x)^2  ]^{\frac{1}{2}}}\\
    f'_2 (x) &= \frac{(1-x)^{2k-1}}{[ 2^{2k}-1 + (1-x)^{2k}  ]^{\frac{2k-1}{2k}}}.
\end{eqs}
For $0< x < 1$ and $k>1$, we have $(1-x) > (1-x)^{2k-1}$ and $ [3 + (1-x)^2  ]^{\frac{1}{2}} < [ 2^{2k}-1 + (1-x)^{2k}  ]^{\frac{2k-1}{2k}}$, so $f'_1 (x) > f'_2 (x)$. Starting from the initial point $f_1(0)=f_2(0)=0$, $f_1(x)$ and $f_2(x)$ cannot intersect in the interval $ 0 < x < 1$. Therefore $a^2-b^2+c^2-d^2 \neq 0$ and the solution for Eq.~\eqref{eq: 4 gates conditions} always exists.

\bibliography{biblio.bib}

\end{document}